\documentclass[onecolumn]{aastex631}
\usepackage{amsmath,amstext}
\usepackage[T1]{fontenc}
\usepackage{apjfonts} 
\usepackage[figure,figure*]{hypcap}
\usepackage{url}
\usepackage{CJKutf8}
\usepackage[version=3]{mhchem}
\usepackage{float}
\usepackage{graphicx}
\usepackage{mathtools}
\usepackage[utf8]{inputenc}
\newtagform{eqn}{(eqn.\;}{)}
\usepackage{subfigure}
\usepackage{amsmath}
\usepackage{natbib}
\usepackage{makecell}

\newcommand{\STELLA}{\texttt{STELLA}}

\newcommand{\MESA}{\texttt{MESA}}
\newcommand{\MESAstar}{\texttt{MESAstar}}
\newcommand{\Ni}{{\ensuremath{^{56}\mathrm{Ni}}}}
\newcommand{\Fe}{{\ensuremath{^{56}\mathrm{Fe}}}}

\newcommand{\Co}{{\ensuremath{^{56}\mathrm{Co}}}}

\newcommand{\Hy}{{\ensuremath{^{1} \mathrm{H}}} }
\newcommand{\Ox}{{\ensuremath{^{16}\mathrm{O}}}}

\newcommand{\Sx}{{\ensuremath{^{32}\mathrm{S}}}}
\newcommand{\Si}{{\ensuremath{^{28}\mathrm{Si}}}}

\newcommand{\Cx}{{\ensuremath{^{12}\mathrm{C}}}}
\newcommand{\Nx}{{\ensuremath{^{14}\mathrm{N}}}}

\newcommand{\foe}{{\ensuremath{\mathrm{foe}}}}
\newcommand{\erg}{{\ensuremath{\mathrm{erg}}}}

\newcommand{\K}{\ensuremath{\mathrm{K}}}

\newcommand{\kms}{\ensuremath{\mathrm{km}\,\mathrm{s}^{-1}}}
\newcommand{\cssa}{\ensuremath{\mathrm{cm}^{-2}}\,\mathrm{s}^{-1}\,$\AA$^{-1}}

\newcommand{\Msun}{\ensuremath{\mathrm{M}_\odot}}
\newcommand{\Rsun}{\ensuremath{\mathrm{R}_\odot}}
\newcommand{\Zsun}{\ensuremath{\mathrm{Z}_{\odot}}}

\shorttitle{Progenitor Stars of Type IIP Supernovae}
\shortauthors{You et al.}

\begin{document}
\begin{CJK*}{UTF8}{bsmi}
\title{Modeling the Progenitor Stars of Observed IIP Supernovae}

\author[0000-0003-3905-1318]{Kai-An You（游凱安）}
\affiliation{Department of Electrical Engineering, National Tsing Hua University, No. 101, Sec. 2, Guangfu Rd., Hsinchu 30013, Taiwan} 

\affiliation{Institute of Astronomy and Astrophysics, Academia Sinica, No.1, Sec. 4, Roosevelt Rd., Taipei 10617, Taiwan}

\correspondingauthor{Kai-An You}
\email{timyu930218@gmail.com}

\author[0000-0002-4848-5508]{Ke-Jung Chen（陳科榮）}
\affiliation{Institute of Astronomy and Astrophysics, Academia Sinica, No.1, Sec. 4, Roosevelt Rd., Taipei 10617, Taiwan} 

\author[0000-0001-8415-6720]{Yen-Chen Pan（潘彥丞）}
\affiliation{Graduate Institute of Astronomy,  National Central University, No.300, Zhongda Rd., Taoyuan 320317, Taiwan} 

\author[0000-0001-5466-8274]{Sung-Han Tsai（蔡松翰）}
\affiliation{Institute of Astronomy and Astrophysics, Academia Sinica, No.1, Sec. 4, Roosevelt Rd., Taipei 10617, Taiwan} 

\affiliation{Department of Physics, National Taiwan University, No.1, Sec. 4, Roosevelt Rd.,  Taipei 10617, Taiwan} 

\author[0000-0003-1295-8235]{Po-Sheng Ou （歐柏昇）}
\affiliation{Institute of Astronomy and Astrophysics, Academia Sinica, No.1, Sec. 4, Roosevelt Rd., Taipei 10617, Taiwan} 
\affiliation{Department of Physics, National Taiwan University, No.1, Sec. 4, Roosevelt Rd.,  Taipei 10617, Taiwan} 

\begin{abstract}
Type IIP supernovae (SNe IIP) are thought to originate from the explosion of massive stars $> 10\;\Msun$. Their luminosity is primarily powered by the explosion energy and the radioactive decay energy of \Co, with the photosphere location regulated by hydrogen recombination. However, the physical connections between SNe IIP and their progenitor stars remain unclear. This paper presents a comprehensive study of SNe IIP and their progenitor stars by using the one-dimensional stellar evolution code, \texttt{MESA}. Our model grids consider the effects of stellar metallicity, mass, and rotation in the evolution of massive stars, as well as explosion energy and \Ni\ production in modeling supernovae. To elucidate the observed SNe IIP and their origins, we compare their light curves (LCs) with our models.
Furthermore, we investigate the impact of stellar parameters on LCs by considering stellar mass metallicity, rotation, explosion energy, and \Ni\ production. We find that more massive stars exhibit longer plateaus due to increased photon diffusion time caused by massive ejecta. Higher metallicity leads to increased opacity and mass loss of progenitor stars. 
Rapid rotation affects internal stellar structures, enhancing convective mixing and mass loss, potentially affecting the plateau's brightness and duration. Higher explosion energy results in brighter but shorter plateaus due to faster-moving ejecta. \Ni\ mass affects late-time luminosity and plateau duration, with larger masses leading to slower declines.
 \end{abstract}

\section{Introduction}

A core-collapse supernova (CCSN) is a dramatic event that occurs when a massive star approaches the end of its life cycle \citep{Woosley&Janka(2005), Fryer&New(2011)}. During the phase of CCSN, an immense amount of radiation is emitted, temporarily outshining the entire galaxy and dispersing newly formed heavy elements into the surrounding interstellar medium. This process plays a crucial role in the evolution of the cosmos. CCSNe are classified into various types based on their light curves (LCs) and spectral features \citep{Filippenko(1997)}. For instance, Type II SNe, which result from the death of massive stars, are characterized by the presence of hydrogen lines in their spectra. Subtypes within Type II SNe are further defined by their LC shapes, such as Type IIP (plateau) and Type IIL (linear) \citep{Valentietal.(2016)}. However, observational studies of well-observed SNe IIP, particularly those with identified progenitor stars, are rare. By carefully examining pre-explosion images from archival data, we can potentially identify the progenitor star of observed SNe and gain insights into the demise of massive stars \citep{Smartt(2009)}.

Previous studies have primarily focused on analyzing various LCs and spectra of SN, using methods like the expanding photosphere (EPM) and standard candle method (SCM) to determine distances and other physical parameters \citep{Leonardetal.(2002)}. These studies have commonly identified the SN types (e.g., Type IIP events), estimated explosion dates, and assessed progenitor characteristics such as stellar mass and age. They have provided insights into the stellar evolution leading to SN explosions, contributing to our understanding of the physical processes involved in these events and refining the technique of cosmological distance estimation \citep{Leonardetal.(2002), Hendryetal.(2005),Vinketal.(2006), Sahuetal.(2006), Taktsetal.(2014), Dall'Oraetal.(2014)}.

To gain deeper insights into the progenitor stars of SNe IIP, we established a grid of stellar evolution models using \MESA\ and simulated their explosions with \STELLA\ to generate multi-color LCs. These models encompass critical parameters for stellar evolution, including mass, metallicity, rotation, mass loss, and factors pertinent to SNe, such as explosion energy and \Ni\ production. We subsequently fit our LC database with those observed in SNe IIP events and deduce the most suitable progenitor models for these SNe. Through this approach, we aim to reveal the physical properties of progenitor stars during their evolution and determine key factors that influence the formation of IIP SNe. Furthermore, we highlight the importance of photometric observations in constraining SN properties and progenitor evolution and attempt to explain the underlying physics of SN IIP emissions.

Section 2 describes our numerical methods of \MESA\ and \STELLA\, followed by the presentation of the stellar evolution models of SN IIP progenitor stars in Section 3. Section 4 presents the results of SN explosions and LC calculations. In Section 5, we compare our LCs with observations, and in Section 6, we discuss the physical properties of progenitor stars. Section 7 discusses the impact of stellar parameters on SN IIP LCs, and finally, we conclude our findings in Section 8.

\section{\MESA\ and \STELLA\ Framework}

\subsection{The \MESA\ Code}
Our stellar evolution models are calculated using Modules for Experiments in Stellar Astrophysics (\MESA), an open-source code suitable for various applications in stellar astrophysics. We utilize the one-dimensional (1D) stellar evolution module \MESAstar\ (version: r15140), which combines essential physical modules such as stellar wind, nuclear-burning network, stellar rotation, hydrodynamics, and opacities for a stellar atmosphere to model the evolution of massive stars \citep{Paxtonetal.(2011), Paxtonetal.(2013), Paxtonetal.(2015), Paxtonetal.(2018), Paxtonetal.(2019)}.

When the massive star approaches its iron-core collapse phase, we employ a thermal bomb to explode it and calculate its observational signatures with \STELLA\ \citep{Blinnikovetal.(1998), Blinnikovetal.(2000), Blinnikovetal.(2006)}. \STELLA\ incorporates multigroup radiative transfer and implicitly solves time-dependent equations for intensity averaged over fixed frequency bands. It is utilized when the shock of the blast wave is near the stellar surface and can simulate shock breakout and interaction with circumstellar medium (CSM) outside the stellar photosphere. It can calculate multi-color LCs and provide observational signatures of SNe IIP. The following subsections will introduce the physics adopted in our \MESA\  and \STELLA\ simulations.

\subsubsection{Rotation}
Stars form through the gravitational collapse in molecular clouds. The inherited angular momentum of the cloud likely leads to the formation of a rotating star. Rotation can significantly impact stellar evolution \citep{Maeder(2009)}. It can induce instabilities that cause angular momentum and chemical elements to be redistributed within the star. When the redistribution of angular momentum causes the star to spin faster, it can increase the centrifugal force, which can cause the star to become oblate. This can lead to a decrease in the effective gravity at the equator, which can affect the star's overall structure and evolution \citep{Hegeretal.(2000)}, including its size, temperature, luminosity, and surface abundance of various elements. Additionally, rotation can influence the star's mass loss rate, further affecting its evolution \citep{Hegeretal.(2000), Paxtonetal.(2013)}.

The effects of rotation become more critical as the star contracts and close-to-rigid rotation is established throughout the star in the phase of zero-age main sequence (ZAMS). The rotation results can be inferred using advanced numerical methods, which can simulate the redistribution of angular momentum and chemical elements within the star and the effects of the centrifugal force on the star's internal structure and mass loss rate. We consider rotation in our models by adjusting the ratio ($\Omega$) of surface angular velocity ($v_{\rm rot}$) to the critical angular velocity ($v_{\rm crit}$) at the star's surface \citep{Langer(1997)}:

\begin{equation}
\label{eqn1}  \Omega\;=\;\dfrac{v_{\rm rot}}{v_{\rm crit}}\;.
\end{equation}

Suppose a star exceeds $v_{\rm crit}$. In that case, it will start to eject material from the surface and disintegrate, causing it to fail to evolve \citep{Hegeretal.(2000), Maeder(2009)}. This phenomenon is known as rotational mass loss and can significantly impact the star's evolution.

\subsubsection{Rotation and Stellar Wind Driven Mass Loss}

Mass loss from stellar winds can significantly affect the evolution of massive stars \citep{Chiosi&Maeder(1986), Ouetal.(2023)}. Besides the stellar wind, rotation can also influence the mass loss rate of stars, further impacting their evolution \citep{Hegeretal.(2003)}. The centrifugal force can cause material to be ejected from the star's equator, increasing the mass loss rate. This can cause the star to lose mass faster than it would otherwise, affecting its evolution \citep{Hegeretal.(2000)}. The mass loss rate also depends on the star's surface rotation rate, with fast-rotating stars losing more mass than slow-rotating stars. This can lead to a feedback loop, where the mass loss rate influences the star's rotation rate, affecting the mass loss rate itself \citep{Ekstrmetal.(2012)}. In the current study, we use the empirical calculation value from \cite{Langer(1997)} (Eqn. \ref{eqn2}), and these values were further modified to explain the effects of stellar rotation \citep{Friend&Abbott(1986), Langer(1997), Heger&Langer(1998), Meynet&Maeder(2000)}:

\begin{equation}
\label{eqn2} {\dot{M}}_{\odot}(\Omega)\;=\;{\dot{M}}_{\odot}(\Omega=0)(\dfrac{1}{1-\Omega})^{\zeta},\quad \zeta\;=\;0.43\;.
\end{equation}
    
In Eqn. \ref{eqn2}, ${\dot{M}}_{\odot}(\Omega=0)$ represents the initial mass loss rate without rotation, which means $v_{\rm rot}\;=\;0$, ${\dot{M}}_{\odot}(\Omega)$ is the rate accounting for rotation, and $\zeta$ is a reference value in the stellar process \citep{Langer(1997)}. If a star reaches the critical angular velocity, it will be torn apart when the centrifugal force exceeds gravity. This equation is expressed in Eqn. \ref{eqn3} \citep{Langer(1997)}. $M_{\odot}$ and $R_{\odot}$ refer to the stellar mass and radius.

In Eqn. \ref{eqn3}, the Eddington luminosity is the maximum luminosity that allows stars to balance radiation outward and gravitational force inward. When a star exceeds the Eddington luminosity, it initiates a strong stellar wind from its outer layers, potentially leading to an explosive event \citep{Maeder(2009)}. Therefore, when a star surpasses this limit, the mass loss process starts to operate until the rotation slows down. Here, $\Gamma$ is the ratio of the stellar to Eddington luminosity, and $L_{\odot}$ and $\kappa$ refer to stellar luminosity and the opacity of the stellar material:

\begin{equation}
\label{eqn3} {v_{\rm crit}}^{2}\;=\;(1-\Gamma)\dfrac{GM_{\odot}}{R_{\odot}}\; ;
\quad \Gamma\;=\;\dfrac{\kappa L_{\odot}}{4\pi c GM_{\odot}}\;.
\end{equation}

The stellar wind is the flow of material emitted from the star's surface, representing a way for the star to lose mass \citep{Hegeretal.(2003)}. The mass loss rate of the stellar wind goes down as the star's mass decreases \citep{Maeder&Meynet(2000)}. For low- and medium-mass stars like the Sun, the mass loss caused by the stellar wind is negligible \citep{Beech(2019)}. However, for massive stars, the mass loss caused by the stellar wind is significant enough to influences their evolution \citep{Maeder&Meynet(2000), Ouetal.(2023)}. In this study, we employ a standard recipe for the stellar wind in \MESA. When the photosphere temperature $<\;0.8\times\;10^{4}\; \;\K$, cool wind schemes are activated; conversely, when the photosphere temperature $>1.2\times\;10^{4}\;\K$, hot wind schemes are activated. 

\subsubsection{Opacity}

In \MESAstar, the application of the grey atmosphere is the primary method to determine the temperature and essential radiation characteristics of astronomical objects (stars, interstellar gases, dust clouds, etc.) \citep{Mihalas(1978)}. We use the local thermal equilibrium (LTE) model to understand temperature behavior as a function of optical depth to infer the interaction between radiative equilibrium and transfer. In Eqn. \ref{eqn6}, $T$ is the internal black body temperature, and $q(\tau)$ is the Hopf function, which is a varying function of optical depth, with only the constant $\dfrac{2}{3}$ replaced.

\begin{equation}
\label{eqn6} T^{4}\;=\;\dfrac{3}{4}{T^{4}_{\rm eff}}[\tau+q(\tau)]\;.
\end{equation}

Optical depth refers to the degree of light absorption as it passes through a medium. The photosphere thickness is defined from the surface to an optical depth of $\dfrac{2}{3}$, at which point the temperature ($T$) equals the effective temperature ($T_{\rm eff}$). At this depth, the energy radiated by the star is equivalent to the total radiant energy observed.

For opacities in LC calculations, \STELLA\ considers various components to accurately model astrophysical conditions. The opacity in \STELLA\ includes photoionization, free-free absorption, and electron scattering, assuming local thermal equilibrium (LTE) in the plasma for ionization and level populations using the Boltzmann-Saha distribution. This is done by considering H, He, heavy elements, stable Fe, radioactive \Co, stable \Ni, and radioactive \Ni. Furthermore, it considers over 153,000 spectral lines from \cite{Kurucz&Bell(1995)} and \cite{Verneretal.(1996)} and employs the expansion opacity rule by \cite{Eastman&Pinto(1993)}.

\subsubsection{Mixing}
The mixing length model is used in astrophysics to describe the convective transport of energy within a star's interior, particularly in regions where the temperature gradient is steep enough to drive convection. In this model, the convective motion is approximated by assuming that material moves in discrete cells, each with a characteristic size known as the mixing length (see Eqn. \ref{eqn7}) \citep{Cox&Giuli(1968)}, which represents the average distance a parcel of material travels before it exchanges energy with its surroundings.

The mixing length model helps predict stellar properties, including surface temperature, luminosity, and evolutionary tracks on the Hertzsprung-Russell (HR) diagram. It is a simplified approach compared to more complex multidimensional simulations but has proven to be quite effective in understanding the overall behavior of stars.

\begin{equation}
\label{eqn7} \Lambda\;=\;\alpha_{\rm MLT}\lambda_{\rm P}\;,
\end{equation}

$\Lambda$ and $\lambda_{P}$ refer to mixing length and pressure scale height, and $\alpha_{\rm MLT}$ is the mixing length parameter (set at 1.5), which is often used to compare the reduction rate of same physical quantity in different environments \citep{Paxtonetal.(2011)}.

\subsubsection{Explosion setup}

The current approach for setting up the explosion of massive stars in \MESAstar\ involves removing the central iron core (approximately mass coordinate $\leq 1.5\;\Msun$) that would have collapsed to form a proto-neutron star. The subsequent explosion is achieved by injecting explosive energy at the outer boundary of the removed iron core \citep{Paxtonetal.(2018)}. Such setups account for gas dynamics and nucleosynthesis during the explosion.

Explosive nucleosynthesis yields in the artificial triggering have significant uncertainties. \MESAstar\ sets the explosion energy at the outer edge of the iron core for a specific duration \citep{Paxtonetal.(2015), Paxtonetal.(2018)}. During this period, the explosion strongly influences the silicon and oxygen-burning shells, with the former primarily contributing to the production of \Ni\ and the latter to the production of \Si\ and \Sx. \MESA's isotope mapping rules involve copying abundances if an isotope exists in both old and new networks, setting absent isotopes to zero mass fraction in the new network, and independently renormalizing all elements to maintain consistent total mass fractions. This method ensures the sum of mass fractions between the old and new networks, providing a reliable and consistent mapping process. Also, the explosion energy deposited rate ($\erg$ $\rm s^{-1}$) significantly impacts abundance profiles, where high deposited rates may yield ten times more \Ni\ formation in the core than those with low deposited rates. Increasing explosion energy or adjusting the location of mass cuts also alters the \Ni\ yields, affecting its LC formation. To mitigate the uncertainty of \Ni\ yield during explosive burning and fallback, we set \Ni\ mass as a parameter in our models.

\subsubsection{Light Curve Calculations with \STELLA}

\STELLA\ is a powerful tool for modeling the LCs of SNe, utilizing an implicitly differenced hydrodynamics approach along with multigroup radiative transfer. It uses the intensity momentum approximation in each frequency bin with 40 to 200 frequency groups to produce bolometric luminosities and broad-band colors \citep{Paxtonetal.(2018)}. The code solves the conservation equations for mass, momentum, and total energy on a Lagrangian comoving grid \citep{Paxtonetal.(2018)}, utilizing artificial viscosity for stability \citep{VonNeumann&Richtmyer(1950)}. The radiation hydrodynamics equations are solved using an implicit high-order predictor-corrector procedure based on the numerical methods \citep{Gear(1971), Blinnikovetal.(1998)}.

The LC calculations unfold in two phases. After core collapse, \MESAstar\ is employed to model the explosion until right before the shock breakout. The resulting profiles of explosion are then mapped on \STELLA\, which further evolve through shock breakout to the post-plateau phase till the fading of SNe IIP. Because the post-explosion evolution is influenced by shock traversal through the stellar interiors that causes a significant mixing of ejecta, e.g., \cite{Chenetal.(2020)}, \cite{Onoetal.(2020)}. \MESA\ considers the multidimensional effects mixing of Rayleigh–Taylor instability by modifying the 1D hydro equations based on \cite{Duffell(2016)}; thus, progenitor parameters such as explosion energy and \Ni\ mass lead to diverse outcomes in modeling LCs \citep{Paxtonetal.(2018)}.

Originally, the energy released from the \Ni\ decays through emitting high-energy gamma rays. Meanwhile, the gamma rays scatter within the dense ejecta and heat the gas to $4,000\;-\;8,000\;\K$. So, the SN emission falls primarily in the visible and infrared wavelengths. \STELLA's gamma-ray transfer computation uses a one-group approximation for nonlocal energy deposition, considering only purely absorptive opacity due to the distinct interaction of gamma-ray photons with ejecta compared to optical/UV photons. Hydrodynamics and multigroup radiative transfer allow the code to depict the SN's luminosity evolution accurately \citep{Blinnikovetal.(1998), Blinnikovetal.(2000)}. Also, the numbers of cells and bins used in STELLA are free parameters, e.g., \cite{Noebaueretal.(2016)}. In this study, we employ 400 Lagrangean zones and 40 frequency bins, enabling an accurate representation of nonequilibrium continuum radiation. The code couples multigroup radiative transfer with hydrodynamics, facilitating a self-consistent calculation of color temperature without additional thermalization depth estimates \citep{Blinnikovetal.(2000)}. 

\STELLA\ computes expansion opacity for each time step in each mesh zone, ensuring reliability in capturing fine details for models with minor parameter differences. 
 Additionally, the code evolves the abundances of iron group elements by considering the decay chain; we only consider the decay chain of \Ni\ → \Co\ → \Fe\ because of its dominating role in light curve calculations of SN IIP \citep{Blinnikovetal.(2006)}. While we acknowledge that the existence of alternative decay chains is also available in STELLA, e.g., \cite{Noebaueretal.(2017)}. To compute variable Eddington factors, \STELLA\ considers scattering and redshifts for each frequency group in each mass zone. The code avoids assumptions about radiative equilibrium, which is crucial during shock breakout. Lastly, it computes the LC by integrating radiation flux at each time step, considering absorption, emission, and scattering within the ejecta. The resulting bolometric and broad-band $UBVRI$ LCs are then compared to observation, providing insights into the SNe's physics and validating the models' accuracy \citep{Blinnikovetal.(2000)}.

\subsection{Model Grid}

We construct a grid of stellar models with varying stellar masses ($13\;\Msun$, $15\;\Msun$, $20\;\Msun$, $25\;\Msun$, $30\;\Msun$, $35\;\Msun$)\footnote{We use the initial stellar mass $>10 \Msun$ to avoid the type I SN progenitors.}, absolute metallicity $(Z\;=\;0.02,\;0.002,\;0.0002)$, and the ratio of angular velocity to the critical angular velocity at the surface ($\Omega$ = 0, 0.1, 0.2, 0.4). These parameters cover the possible progenitor stars of SN IIP across cosmic time, and we present 72 models before the explosion phase. When the massive stars evolve to the iron core collapse, we explode them with explosion energies $(0.6\;\foe,\;1.2\;\foe,\;1.8\;\foe,\;2.4\;\foe,\;1\;\foe = 10^{51}\;\erg)$ with nickel mass $M_{\rm Ni}\;=\;0.01\;\Msun,\;0.02\;\Msun,\;0.05\;\Msun,\;0.1\;\Msun)$. We follow the explosion for 200 days and calculate their LCs in \STELLA\ with 1152 models in total \footnote{All inlist files for reproducing our \MESA\ models are available at \dataset[10.5281/zenodo.11172532]{\doi{10.5281/zenodo.11172532}}.}. After obtaining a large grid of LCs, we compare these LCs with SN IIP observations. These data are shown in section 2.3. 

\subsection{Data of Observed SNe IIP}

\begin{table*}[tbh]
\centering
\hskip-2.2cm
\resizebox{20.2cm}{!}{%
\begin{tabular}{|l|cccccccccc|lll|}
\hline
SN           & Host      & \hspace{-2em}\begin{tabular}[c]{@{}c@{}}Explosion date\\ (MJD)\end{tabular} & R.A. (J2000)                     & Dec. (J2000)                     & $v_{r}$ (km/s) & d (Mpc)                             & $E(B\;-\;V)$ & \hspace{-2em}\begin{tabular}[c]{@{}c@{}}Flux\\ ($10^{-9}\;\erg\;\cssa$)\end{tabular} & \hspace{-2em}\begin{tabular}[c]{@{}c@{}}$v_{\rm ph}$\\ (\kms)\end{tabular} & \hspace{-2em}\begin{tabular}[c]{@{}c@{}}Plateau duration\\ (day)\end{tabular} & \multicolumn{3}{l|}{Reference}                    \\ \hline
$SN\;1999em$ & NGC 1637  & 51480                                                          & 04:41:27.04                      & -02:51:45.20                     & 710            & $7.7\;\pm\;0.4$                     & 0.036        & 0.01508                                                               & $\sim3500$                                                & 128.8                                                            & \multicolumn{3}{l|}{\cite{Leonardetal.(2002)}}    \\
$SN\;1999gi$ & NGC 3184  & 51521                                                          & 10:18:16.66                      & +41:26:28.20                     & 588            & $11\;\pm\;1$                        & 0.014        & 0.01080                                                               & $6374\;\pm\;144$                                                         & 130.4                                                            & \multicolumn{3}{l|}{\cite{Leonardetal.(2002)-1}}  \\
$SN\;2003gd$ & NGC 628   & 52715                                                          & \multicolumn{1}{l}{01:36:42.650} & \multicolumn{1}{l}{+15:44:20.90} & 656            & \multicolumn{1}{l}{$9.3\;\pm\;1.8$} & 0.061        & 0.01290                                                               & -                                                         & 128                                                              & \multicolumn{3}{l|}{\cite{Hendryetal.(2005)}}     \\
$SN\;2004A$  & NGC 6207  & 53014                                                          & 16:43:01.90                      & +36:50:12.50                     & 853            & $20\;\pm\;3$                        & 0.013        & 0.00764                                                               & $4744\;\pm\;29$                                                         & 130.49                                                           & \multicolumn{3}{l|}{\cite{Tsvetkov(2008)}}        \\
$SN\;2004dj$ & NGC 2403  & 53217                                                          & 07:37:17.02                      & +65:35:57.80                     & 130            & $3.5\;\pm\;0.5$                     & 0.058        & 0.03099                                                               & $\sim4350$                                          & $\sim120$                                                        & \multicolumn{3}{l|}{\cite{Vinketal.(2006)}}       \\
$SN\;2004et$ & NGC 6946  & 53275                                                          & 12:21:54.89                      & +04:28:25.30                     & 1574           & $14\;\pm\;2$                        & 0.020        & 0.02480                                                               & $\sim5060$                                                & $\sim143$                                                        & \multicolumn{3}{l|}{\cite{Sahuetal.(2006)}}       \\
$SN\;2005cs$ & M51       & 53549                                                          & 13:29:52.850                     & +47:10:36.30                     & -              & $7.1\;\pm\;1.2$                     & 0.031        & 0.01137                                                               & $\sim1700$                                                & 129.64                                                           & \multicolumn{3}{l|}{\cite{Pastorelloetal.(2009)}} \\
$SN\;2007od$ & UGC 12846 & 54404                                                          & 23:55:48.680                     & +18:24:54.80                     & 1734           & 24.5                                & 0.033        & 0.01309                                                               & $\sim5417$                                                & $\sim100$                                                                & \multicolumn{3}{l|}{\cite{Inserraetal.(2011)}}    \\
$SN\;2008in$ & NGC 4303  & 54826                                                          & 12:22:01.77                      & +04:28:47.50                     & 1574           & $14\;\pm\;2$                        & 0.020        & 0.00815                                                               & $\sim3250$                                                & 115                                                              & \multicolumn{3}{l|}{\cite{Royetal.(2011)}}        \\
$SN\;2009bw$ & UGC 2890  & 54916                                                          & 03:56:02.92                      & +72:50:42.90                     & 1155           & 20.0                                & 0.198        & 0.00975                                                               & $\sim3050$                                          & 138.4                                                            & \multicolumn{3}{l|}{\cite{Inserraetal.(2012)}}    \\
$SN\;2009js$ & NGC 918   & 55115                                                          & 02:25:48.281                     & +18:29:25.80                     & -              & $21.7\;\pm\;1.8$                    & 0.315        & 0.00373                                                               & $3728\;\pm\;35$                                                         & 109.1                                                            & \multicolumn{3}{l|}{\cite{Gandhietal.(2013)}}     \\
$SN\;2009N$  & NGC 4487  & 54855                                                          & 12:31:09.46                      & -08:02:56.30                     & 1091           & $18\;\pm\;2$                        & 0.019        & 0.00564                                                               & $\sim2250$                                                & 110.5                                                            & \multicolumn{3}{l|}{\cite{Taktsetal.(2014)}}      \\
$SN\;2012A$  & NGC 3239  & 55933                                                          & 10:25:07.39                      & +17:09:14.60                     & 830            & 8.82                                & 0.029        & 0.01484                                                               & $\sim3600$                                                & 103.75                                                           & \multicolumn{3}{l|}{\cite{Tomasellaetal.(2013)}}  \\
$SN\;2012aw$ & NGC 3351  & 56002                                                          & 10:43:53.76                      & +11:40:17.90                     & 771            & 11.507                              & 0.025        & 0.01850                                                               & $\sim4900$                                                         & $\sim140$                                                                & \multicolumn{3}{l|}{\cite{Dall'Oraetal.(2014)}}   \\
$SN\;2012ec$ & NGC 1084  & 56150                                                          & 02:45:59.89                      & -07:34:25.00                     & 1414           & 20.844                              & 0.023        & 0.01003                                                               & $\sim4575$                                          & 109.08                                                           & \multicolumn{3}{l|}{\cite{Barbarinoetal.(2015)}}  \\
$SN\;2013ab$ & NGC 5669  & 56340                                                          & 14:32:44.49                      & +09:53:12.30                     & 1338           & 23.64                               & 0.025        & 0.01052                                                               & $\sim4500$                                                & 105.73                                                           & \multicolumn{3}{l|}{\cite{Boseetal.(2015)}}       \\
$SN\;2013ej$ & M74       & 56497                                                          & 01:36:46.16                      & +15:35:41.00                     & -              & $9.6\;\pm\;0.7$                     & 0.057        & 0.02547                                                               & $\sim5825$                                                & 104.25                                                           & \multicolumn{3}{l|}{\cite{Huangetal.(2015)}}      \\ \hline
\end{tabular}%
}
\caption{Information of well-observed SNe IIP. We determine the photospheric velocity (\(v_{\rm ph}\)) around 50 days after the star explodes and calculate \(v_{\rm ph}\) by averaging velocities measured from the emission lines of hydrogen and iron. For SNe without \(v_{\rm ph}\) information in their references, we retrieve \(v_{\rm ph}\) using the analysis method from \cite{Panetal.(2022)}.}
\label{tab1:CCSN info}
\end{table*}

We use $The\;Open\;Astronomy\;Catalogs$ to retrieve the observational data of SNe IIP in the literature \footnote{Website from \url{https://github.com/astrocatalogs/OACAPI}.}. We transform the apparent magnitude ($m$) of observed SNe to the absolute magnitude ($M$) by considering the effect of Milky Way Extinction. We examine all well observed SNe IIP, including  $SN\;1999em$, $SN\;1999gi$, $SN\;2003gd$, $SN\;2004A$, $SN\;2004dj$, $SN\;2004et$, $SN\;2005cs$, $SN\;2007od$, $SN\;2008in$, $SN\;2009bw$, $SN\;2009N$, $SN\;2012A$, $SN\;2012aw$, $SN\;2012ec$, $SN\;2013ab$, and $SN\;2013ej$ and list all of them in Table \ref{tab1:CCSN info}. We choose Landolt $UBVRI$ as our examined wavelengths ($U$: $0.3508\;\rm µm$, $B$: $0.4329\;\rm µm$, $V$: $0.5422\;\rm µm$, $R$: $0.6428\;\rm µm$, $I$: $0.8048\;\rm µm$). Color indexes of observed SNe have been corrected for the extinction effect of the Milky Way and the host galaxy with an extinction calculator of the equatorial system from $NASA/IPAC\;Extragalatic\;Database$ \footnote{Website from \url{http://ned.ipac.caltech.edu/extinction_calculator} by \cite{Schlafly&Finkbeiner(2011)}.}: 

\begin{equation}
\label{eqn9} M\;=\;m\;-\;5(\log_{10}pc\;-\;1)\;-\;A_{\lambda}\;,
\end{equation}

where $pc$ is the distance between the Earth and the object in parsecs, $A_{\lambda}$ is galactic extinction. In the paper of \cite{Schlafly&Finkbeiner(2011)}, the authors estimate host galaxy extinction by measuring the difference between the observed colors of stars with spectra in the Sloan Digital Sky Survey (SDSS) and the predicted intrinsic colors derived from stellar parameters. This work recommended a $14\%$ recalibration of the SFD dust map since $E(B\;-\;V)\;=\;0.86$ \citep{Schlegeletal.(1998)}. They utilize the Sloan Extension for Galactic Understanding (SEGU) and Exploration Stellar Parameter Pipeline (SSPP) to obtain stellar parameters and connect them to colors using the Model Atmospheres with a Radiative and Convective Scheme (MARCS grid) \citep{Gustafssonetal.(2008)}, which is 1D, hydrostatic, plane-parallel and spherical LTE model atmospheres. By comparing the observed and predicted colors, they calculate the reddening for each star, representing the extinction due to dust in the host galaxy. The authors use \cite{Fitzpatrick(1999)} reddening law to estimate Milky Way extinction. They find that this law with an $R_{v}\;=\;3.1$ provides a more accurate representation of the extinction affecting high-latitude stars with $E(B\;-\;V)<1$ compared to other reddening laws.

\section{Evolution and Explosion of 13 - 35 \Msun\ Stars}

\begin{table*}[tbh]
\hskip-2.2cm
\resizebox{20.2cm}{!}{%
\centering
\begin{tabular}{|lccc|lccc|lccc|lccc|}
\hline
\multicolumn{4}{|c|}{\textbf{$\Omega\;=\;0$}}                                                                                       & \multicolumn{4}{c|}{\textbf{$\Omega\;=\;0.1$}}                                                                                     & \multicolumn{4}{c|}{\textbf{$\Omega\;=\;0.2$}}                                                                                     & \multicolumn{4}{c|}{\textbf{$\Omega\;=\;0.4$}}                                                                                     \\ \hline
\multicolumn{1}{|l|}{$Metallicity$} & \multicolumn{1}{l}{$Z\;=\;0.02$} & \multicolumn{1}{l}{$Z\;=\;0.002$} & \multicolumn{1}{l|}{$Z\;=\;0.0002$} & \multicolumn{1}{l|}{$Metallicity$} & \multicolumn{1}{l}{$Z\;=\;0.02$} & \multicolumn{1}{l}{$Z\;=\;0.002$} & \multicolumn{1}{l|}{$Z\;=\;0.0002$} & \multicolumn{1}{l|}{$Metallicity$} & \multicolumn{1}{l}{$Z\;=\;0.02$} & \multicolumn{1}{l}{$Z\;=\;0.002$} & \multicolumn{1}{l|}{$Z\;=\;0.0002$} & \multicolumn{1}{l|}{$Metallicity$} & \multicolumn{1}{l}{$Z\;=\;0.02$} & \multicolumn{1}{l}{$Z\;=\;0.002$} & \multicolumn{1}{l|}{$Z\;=\;0.0002$} \\ \hline
\multicolumn{1}{|l|}{$13\;\Msun$}      & 11.6308                    & 12.6206                     & 12.9178                       & \multicolumn{1}{l|}{$13\;\Msun$}      & 11.6597                    & 12.2036                     & 12.9102                       & \multicolumn{1}{l|}{$13\;\Msun$}      & 11.6075                    & 12.4663                     & 12.8874                       & \multicolumn{1}{l|}{$13\;\Msun$}      & 11.5381                    & -                           & 12.8584                       \\
\multicolumn{1}{|l|}{$15\;\Msun$}      & 12.7426                    & 13.0679                     & 14.8741                       & \multicolumn{1}{l|}{$15\;\Msun$}      & 12.7426                    & 13.0679                     & 14.8741                       & \multicolumn{1}{l|}{$15\;\Msun$}      & 12.7450                    & 13.9543                     & 12.8643                       & \multicolumn{1}{l|}{$15\;\Msun$}      & 12.6655                    & 14.3070                     & 14.7652                       \\
\multicolumn{1}{|l|}{$20\;\Msun$}      & 15.7600                    & 19.3757                     & 19.9265                       & \multicolumn{1}{l|}{$20\;\Msun$}      & 17.2504                    & 18.6322                     & 19.9110                       & \multicolumn{1}{l|}{$20\;\Msun$}      & 14.4513                    & 18.2083                     & 19.8725                       & \multicolumn{1}{l|}{$20\;\Msun$}      & 14.1927                    & 14.5550                     & 19.8725                       \\
\multicolumn{1}{|l|}{$25\;\Msun$}      & 15.4891                    & 22.8631                     & 24.8266                       & \multicolumn{1}{l|}{$25\;\Msun$}      & 15.4891                    & 22.8631                     & 24.8266                       & \multicolumn{1}{l|}{$25\;\Msun$}      & 13.6765                    & 22.9477                     & 24.6457                       & \multicolumn{1}{l|}{$25\;\Msun$}      & 12.4781                    & 16.5929                     & -                             \\
\multicolumn{1}{|l|}{$30\;\Msun$}      & 17.9506                    & 26.5449                     & 29.6460                       & \multicolumn{1}{l|}{$30\;\Msun$}      & 17.9506                    & 26.5449                     & 29.6460                       & \multicolumn{1}{l|}{$30\;\Msun$}      & 16.3528                    & 25.2598                     & 19.6288                       & \multicolumn{1}{l|}{$30\;\Msun$}      & 14.0778                    & 20.3983                     & -                             \\
\multicolumn{1}{|l|}{$35\;\Msun$}      & 16.8458                    & 30.0005                     & 32.9872                       & \multicolumn{1}{l|}{$35\;\Msun$}      & 16.8458                    & 30.0005                     & 32.9872                       & \multicolumn{1}{l|}{$35\;\Msun$}      & 17.0144                    & 30.1118                     & 34.0413                       & \multicolumn{1}{l|}{$35\;\Msun$}      & -                          & -                           & -                             \\ \hline
\end{tabular}%
}
\caption{The final remaining stellar masses before the explosion. The blank is where the models fail to evolve to core collapse and will be discussed in Section \ref{sec:modeling}.}
\label{tab2:mass loss}
\end{table*}

\begin{table*}[tbh]
\hskip-2.2cm
\resizebox{20.2cm}{!}{%
\centering
\begin{tabular}{|lccc|lccc|lccc|}
\hline
\multicolumn{4}{|c|}{\textbf{$\Omega\;=\;0.1$}}                                                                                   & \multicolumn{4}{c|}{\textbf{$\Omega\;=\;0.2$}}                                                                                   & \multicolumn{4}{c|}{\textbf{$\Omega\;=\;0.4$}}                                                                                   \\ \hline
\multicolumn{1}{|l|}{$Metallicity$} & \multicolumn{1}{l}{$Z\;=\;0.02$} & \multicolumn{1}{l}{$Z\;=\;0.002$} & \multicolumn{1}{l|}{$Z\;=\;0.0002$} & \multicolumn{1}{l|}{$Metallicity$} & \multicolumn{1}{l}{$Z\;=\;0.02$} & \multicolumn{1}{l}{$Z\;=\;0.002$} & \multicolumn{1}{l|}{$Z\;=\;0.0002$} & \multicolumn{1}{l|}{$Metallicity$} & \multicolumn{1}{l}{$Z\;=\;0.02$} & \multicolumn{1}{l}{$Z\;=\;0.002$} & \multicolumn{1}{l|}{$Z\;=\;0.0002$} \\ \hline
\multicolumn{1}{|l|}{$13\;\Msun$}      & 0.799\%                    & -3.169\%                    & -0.029\%                      & \multicolumn{1}{l|}{$13\;\Msun$}      & 0.997\%                    & -0.992\%                    & -0.172\%                      & \multicolumn{1}{l|}{$13\;\Msun$}      & 2.158\%                    & -                           & -0.304\%                      \\
\multicolumn{1}{|l|}{$15\;\Msun$}      & 0.828\%                    & 0.690\%                     & 0.039\%                       & \multicolumn{1}{l|}{$15\;\Msun$}      & 1.835\%                    & 8.397\%                     & 0.020\%                       & \multicolumn{1}{l|}{$15\;\Msun$}      & 3.917\%                    & 13.608\%                    & -0.525\%                      \\
\multicolumn{1}{|l|}{$20\;\Msun$}      & 10.839\%                   & -3.693\%                    & -0.061\%                      & \multicolumn{1}{l|}{$20\;\Msun$}      & -5.750\%                   & -5.719\%                    & -0.234\%                      & \multicolumn{1}{l|}{$20\;\Msun$}      & -3.572\%                   & -24.281\%                   & -0.180\%                      \\
\multicolumn{1}{|l|}{$25\;\Msun$}      & 2.929\%                    & 0.435\%                     & 0.032\%                       & \multicolumn{1}{l|}{$25\;\Msun$}      & -5.882\%                   & 1.324\%                     & -0.659\%                      & \multicolumn{1}{l|}{$25\;\Msun$}      & -5.129\%                   & -25.720\%                   & -                             \\
\multicolumn{1}{|l|}{$30\;\Msun$}      & 3.211\%                    & 0.607\%                     & 0.055\%                       & \multicolumn{1}{l|}{$30\;\Msun$}      & -2.297\%                   & -3.577\%                    & 0.062\%                       & \multicolumn{1}{l|}{$30\;\Msun$}      & -6.090\%                   & -20.617\%                   & -                             \\
\multicolumn{1}{|l|}{$35\;\Msun$}      & 5.257\%                    & 0.778\%                     & 0.284\%                       & \multicolumn{1}{l|}{$35\;\Msun$}      & 13.296\%                   & 2.084\%                     & 4.766\%                       & \multicolumn{1}{l|}{$35\;\Msun$}      & -                          & -                            & -                             \\ \hline
\end{tabular}%
}
\caption{The deviation ratio between the simulated and theoretical mass of rotating stars at the end of their evolution. The theoretical masses of stars with $\Omega\;=\;0.1\;-\;0.4$ are estimated from Eqn. \ref{eqn3} with the results of the corresponding non-rotating star. The deviation ratio is defined as $[\dfrac{\rm simulated\; final\; mass\;-\;theoretical\;final\;mass}{\rm theoretical\;final\;mass}]\times 100\%$.}
\label{tab3:mass loss with prediction}
\end{table*}

Beginning with the ZAMS phase, the star starts to burn the hydrogen envelope in the main sequence (MS), finally forming neutron stars or black holes. The contraction may pause in this process as nuclear fusion supplies the energy necessary to replenish the star's lost radiation and neutrinos \citep{Bally&Reipurth(2006)}. Whenever an element is burnt completely, the core shrinks because it cannot support its gravity, causing the core density and temperature to increase, reaching the ignition point of the next element and continuing to burn \citep{Woosleyetal.(2002), Woosley&Janka(2005)}.

As the core's temperature approaches $10^{9}$K or higher, silicon and other elements release a proton or alpha particle through photodisintegration \citep{Woosleyetal.(1973), Rauscheretal.(2002)}, the fusion of alpha particles forms into heavier elements through the alpha-catching process. On the other hand, electron-positron annihilation produces neutrino–antineutrino pairs \citep{Woosley&Janka(2005)}. These neutrinos will easily escape the star and force it to burn faster to compensate for the loss. After the silicon burning, \Fe\ is accumulated in the core \citep{Fewell(1995)}, then undergoes a catastrophic collapse and explodes into a CCSN \citep{Woosley&Janka(2005), Srinivasan(2014)}.

\subsection{Final Stellar Mass and Radius}

\begin{figure}[H]
\centering
\includegraphics[scale=0.5]{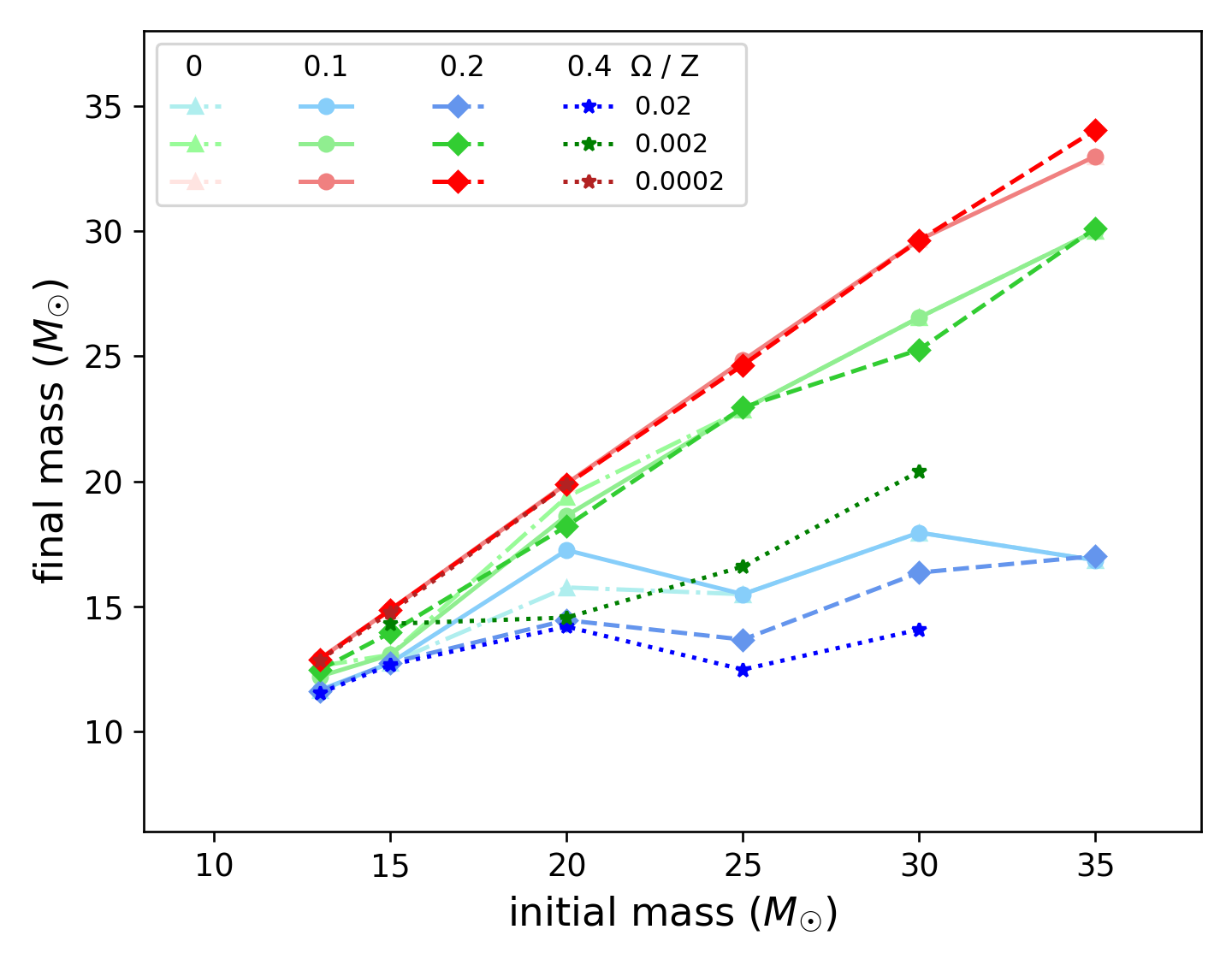}
\caption{The final stellar mass as a function of its initial mass, metallicity, and rotation. Line colors and shapes present various $Z$ and $\Omega$. The figure suggests that this mass function of those low-metallicity stars ($Z\;=0.002,\;0.0002$) positively correlates with the initial mass despite a subtle effect of $\Omega$.}
\label{fig3:mass loss}
\end{figure}

Figure \ref{fig3:mass loss} and Table \ref{tab2:mass loss} show the mass loss of SN IIP progenitor stars associated with their masses, metallicities, and rotations. Generally, the mass loss increases as the star's metallicity and rotation ratio increases. The detail of the metallicity effect on massive stars' evolution can be found in, e.g., \cite{Ouetal.(2023)}, \cite{Cervio&Mas-Hesse(1994)}, and the physics of mass loss driven by rotation are discussed in \cite{Blinnikovetal.(2000)}.

We find that in the case of ($\Omega\;=\;0.4\;/\;Z\;=\;0.002$) for $20\;-30\;\Msun$ stars, the final simulated remaining mass is much lower than the theoretical value base on the case in $\Omega\;=\;0$ we calculate from Eqn. \ref{eqn2}, the deviation is all larger than $20\%$ (see Table \ref{tab3:mass loss with prediction}). The reason is that rapidly rotating models lead to a strong dynamic impact that dredges heavy elements such as \Cx, \Nx, and \Ox\ to the surface during helium-burning. This process substantially raises surface metallicity, ultimately intensifying radiative mass loss through increasing the opacities in stellar atmosphere \citep{Aryanetal.(2023)}.

The final radius before the stars become SNe can affect the observational properties of SNe. We show the final radius as a function of initial mass, metallicity, and rotation in Figure \ref{fig13:star radius} and summarize them in Table \ref{tab4:star radius}. This radius is more strongly correlated with the metallicity, with low-metallicity stars tending to have a smaller final radius. This implies that low-metallicity stars can lead to blue supergiants (BSGs) instead of red supergiants (RSGs) before exploding, which is consistent with \cite{Ouetal.(2023)}. The final radius determines the size of the envelope and can affect the SN's LCs during the shock breakout and plateau phase. The duration of breakout and plateau light curves are longer for stars with a larger final radius.

\subsection{Core Properties}

\begin{figure*}[tbh]
    \centering
    \includegraphics[scale=0.45]{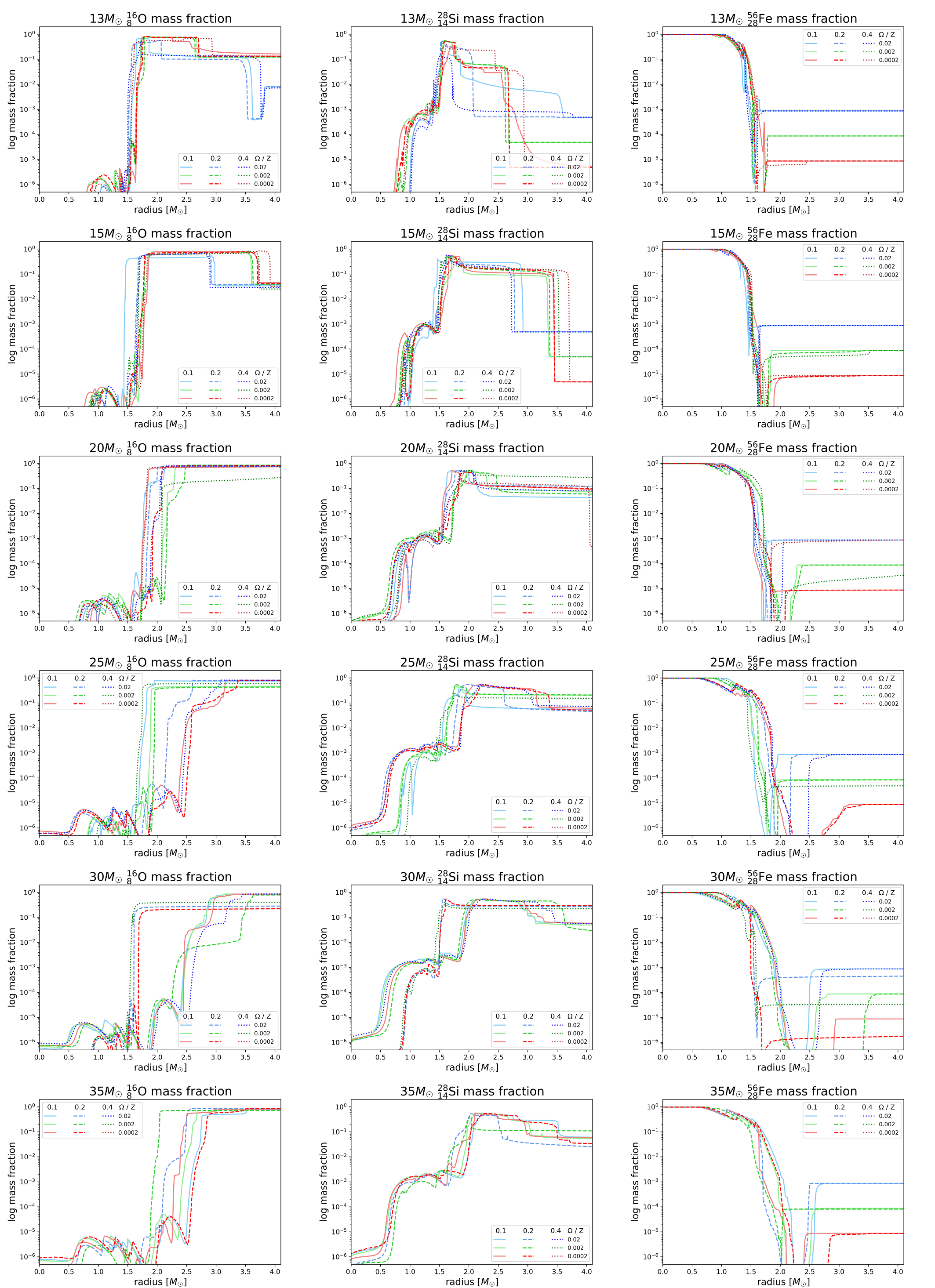}
    \caption{Chemical abundance of \Ox, \Si, and \Fe\ of $13\;-\;35\;\Msun$ stars before core-collapse. Line colors and styles represent various metallicity and rotation, respectively.}
    \label{fig4:abn}
\end{figure*}

\begin{figure}[ht]
    \centering
    	\includegraphics[scale=0.5]{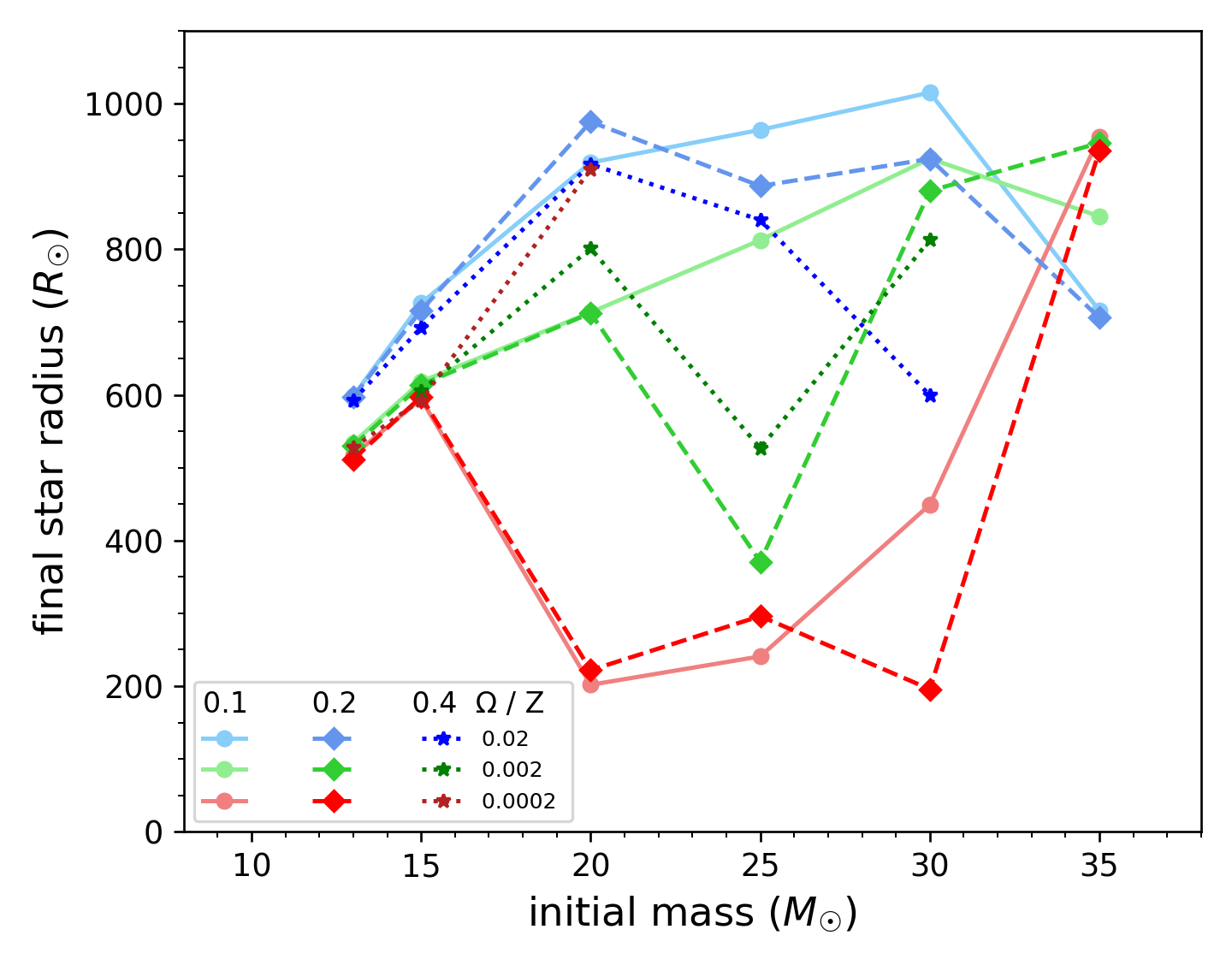}
    \caption{Pre-supernova radius for $13\;-\;35\;\Msun$ stars with different $Z$ and $\Omega$. Generally, the pre-supernova radius seems to have no relation with mass and rotation. The effect of metallicity is more prominent as the higher metallicity can lead to a larger radius.}
    \label{fig13:star radius}
\end{figure}

Figure \ref{fig4:abn} shows the core abundance patterns of \Ox, \Si, \Fe\ in $13\;-\;35\;\Msun$ stars before they die. $Z$ and $\Omega$ can significantly alter the core elemental abundance pattern. Above the iron-rich core of mass coordinate $>1.5\;\Msun$, higher $Z$ tends to deplete more \Ox\ and \Si. Meanwhile, the $\Omega$ also impacts the elemental distribution in the core, but its effect is weaker than $Z$. Based on \Fe\ abundance, $Z$ and $\Omega$ also impact the final iron core mass, which ranges from $1.5\;\Msun$ to $2.3\;\Msun$ depending on the progenitor star mass. The star forms an iron core at the end of the silicon burning. In most cases, the iron core rapidly builds up to $\sim 1.5\;\Msun$. Higher $Z$ causes a difference in distribution around mass coordinates $>2.0\;\Msun$, and \Fe\ is generated more outside the core. Still, the influence of $\Omega$ on the distribution is subtle.

\subsection{Explosions and Shock Breakout}

When the iron core mass reaches the Chandrasekhar mass of $\sim 1.5\;\Msun$, the core then collapses. During the collapse, the photodisintegration dissolves the heavy elements into protons, and the consequent $\beta^{+}$ decay $(p \rightarrow n+e^{+}+v_{e})$ forms a proto-neutron star at the center. When the strong force from the degenerate neutrons stops the runaway collapse, the inwardly collapsing matter in the core bounces up suddenly. The bounce creates an outgoing shock that encounters the infalling material, and then the shock stalled. The copious neutrinos emitted from the proto-neutron star revive the shock that explodes the star \citep{Woosley&Janka(2005), Srinivasan(2014)}. 

\begin{table*}[tbh]
\hskip-2.2cm
\resizebox{20.2cm}{!}{%
\begin{tabular}{|lccc|lccc|lccc|}
\hline
\multicolumn{4}{|c|}{\textbf{$\Omega\;=\;0.1$}}                                     & \multicolumn{4}{c|}{\textbf{$\Omega\;=\;0.2$}}                                     & \multicolumn{4}{c|}{\textbf{$\Omega\;=\;0.4$}}                                     \\ \hline
\multicolumn{1}{|c|}{Metallicity}   & $Z\;=\;0.02$ & $Z\;=\;0.002$ & $Z\;=\;0.0002$ & \multicolumn{1}{l|}{Metallicity}   & $Z\;=\;0.02$ & $Z\;=\;0.002$ & $Z\;=\;0.0002$ & \multicolumn{1}{l|}{Metallicity}   & $Z\;=\;0.02$ & $Z\;=\;0.002$ & $Z\;=\;0.0002$ \\ \hline
\multicolumn{1}{|l|}{$13\;\Msun$} & 597.853      & 534.186       & 516.833        & \multicolumn{1}{l|}{$13\;\Msun$} & 597.508      & 529.747       & 511.372        & \multicolumn{1}{l|}{$13\;\Msun$} & 592.496      & -             & 527.299        \\
\multicolumn{1}{|l|}{$15\;\Msun$} & 725.841      & 618.253       & 594.390        & \multicolumn{1}{l|}{$15\;\Msun$} & 715.791      & 613.670       & 597.400        & \multicolumn{1}{l|}{$15\;\Msun$} & 692.390      & 605.160       & 592.840         \\
\multicolumn{1}{|l|}{$20\;\Msun$} & 919.098      & 713.306       & 201.733        & \multicolumn{1}{l|}{$20\;\Msun$} & 975.044      & 712.305       & 221.905        & \multicolumn{1}{l|}{$20\;\Msun$} & 916.822      & 801.201       & 909.692        \\
\multicolumn{1}{|l|}{$25\;\Msun$} & 963.934      & 812.509       & 240.763        & \multicolumn{1}{l|}{$25\;\Msun$} & 886.825      & 370.814       & 296.067        & \multicolumn{1}{l|}{$25\;\Msun$} & 840.197      & 526.668       & -              \\
\multicolumn{1}{|l|}{$30\;\Msun$} & 1015.165     & 924.421       & 449.231        & \multicolumn{1}{l|}{$30\;\Msun$} & 924.076      & 880.116       & 195.758        & \multicolumn{1}{l|}{$30\;\Msun$} & 599.329      & 812.831       & -              \\
\multicolumn{1}{|l|}{$35\;\Msun$} & 715.472      & 845.181       & 954.333        & \multicolumn{1}{l|}{$35\;\Msun$} & 705.962      & 946.072       & 935.621        & \multicolumn{1}{l|}{$35\;\Msun$} & -            & -             & -              \\ \hline
\end{tabular}%
}
\caption{Pre-supernova radius of $13\;-\;35\;\Msun$ stars of various $\Omega$ and $Z$. The unit of radius is \Rsun.}
\label{tab4:star radius}
\end{table*}

\begin{figure}[h]
    \centering
    \subfigure[]{
        \includegraphics[scale=0.48]{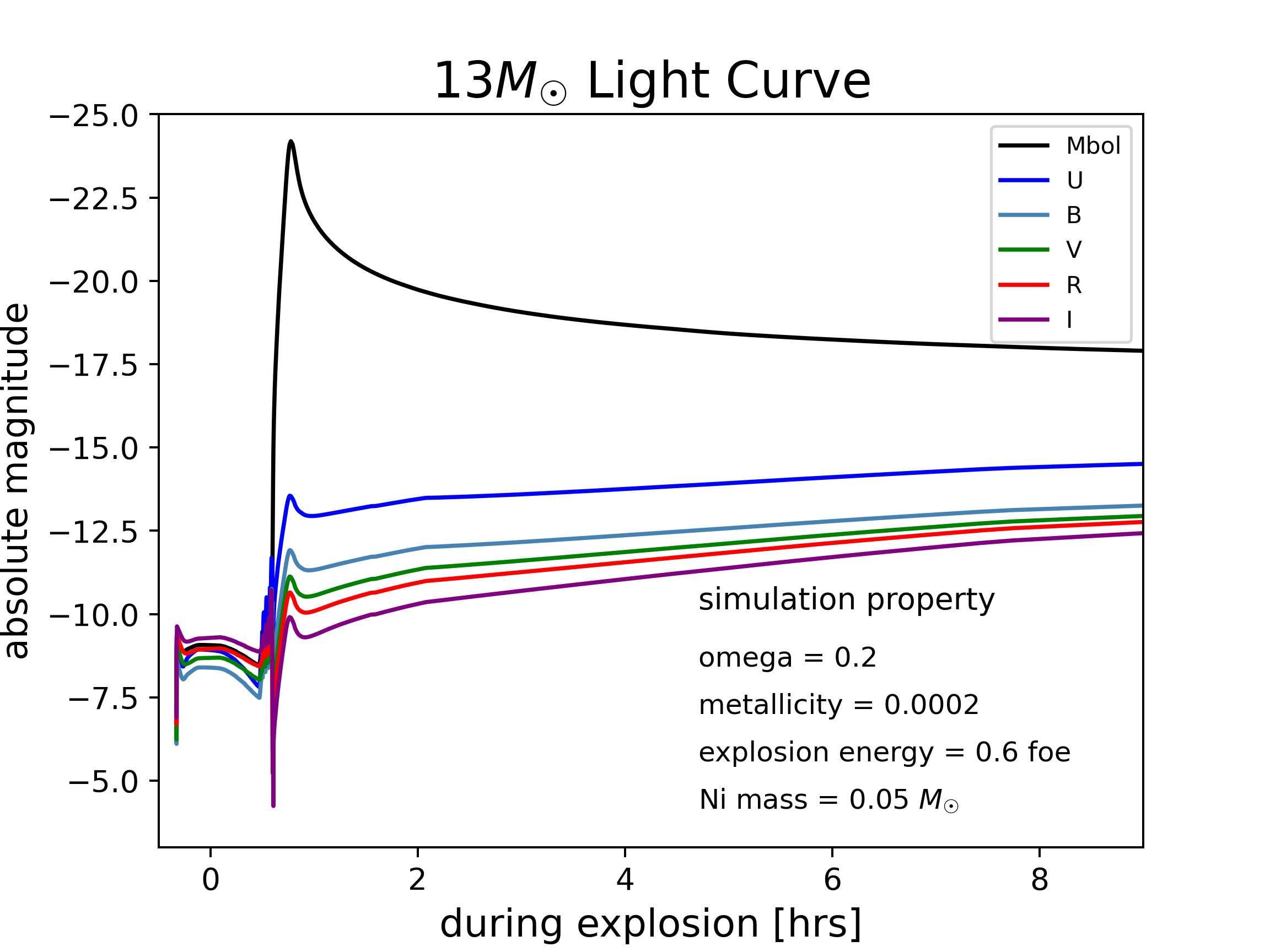}
        }
    \subfigure[]{
        \includegraphics[scale=0.48]{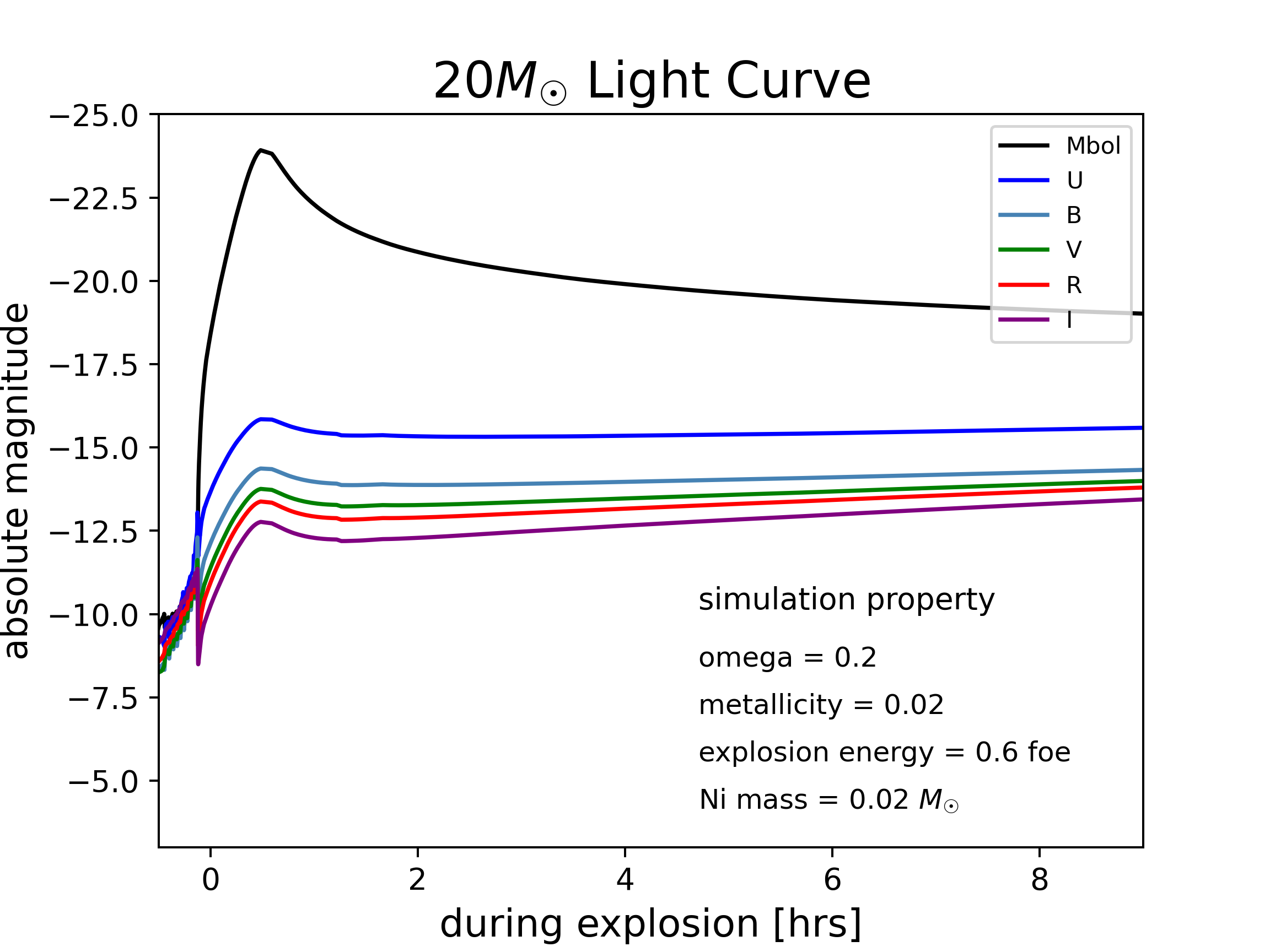}
        }
    \caption{LCs of SN's shock breakout. We show the case (a) $13\;\Msun\;/\;\Omega\;=\;0.2\;/\;Z\;=\;0.0002\;/\;0.6\;\foe\;/\;M_{\rm Ni}\;=\;0.05\;\Msun$ and (b) $20\;\Msun\;/\;\Omega\;=\;0.2\;/\;Z=\;0.02\;/\;0.6\;\foe\;/\;M_{\rm Ni}\;=\;0.02\;\Msun$. The bolometric luminosity of SNe can reach peak within one hour, and its $M_{\rm bol}$ is up to -24.}
    \label{fig17:shock breakout}
\end{figure}

Through the driving mechanism of neutrinos, a part of the potential energy lost in the collapse is transferred to the star's outer layer through neutrino heating, which will trigger a radiation-mediated shock wave at a speed of 8,000 - 10,000 \kms \citep{Jankaetal.(2016)}. The revived shock propagates through the stellar interiors, eventually breaking out the star's surface, and the phenomenon is called shock breakout. The plasma in front of the shock wave can emit electromagnetic radiation to produce a breakout of radiation when its optical depth $\tau \approx\;\dfrac{c}{v_{\rm sh}}$\citep{Ohyama(1963), Srinivasan(2014), Waxman&Katz(2017)}, where $c$ is the speed of light, and $v_{\rm sh}$ is the shock velocity. Once the shock reaches the star's surface, it flushes bright ultraviolet and X-rays for seconds to a couple of hours. Then comes the optical cooling of the ultraviolet light from expansion cooling, which can last from hours to days depending on the shock's extent \citep{Klein&Chevalier(1978), Waxman&Katz(2017)}. If the CSM is formed before the star dies and its optical depth is larger than $\dfrac{c}{v_{\rm sh}}$, the breakout will occur at a larger radius in the CSM with a time scale up to a few days.  Therefore, the properties of the burst are related to the progenitor star's structure (radius and surface composition) and its mass loss history. \citep{Falk&Arnett(1973), Falk&Arnett(1977), Chevalier&Fransson(1994), Chenetal.(2024)}.

In the RSG phase, the shock wave from the explosion is expected to propagate for about a day until it reaches an optical depth for breakout \citep{Frsteretal.(2018)}, and the explosion energy and envelope mass determine the breakout duration \citep{Lovegroveetal.(2017)}. We present two representative shock breakouts from 13 and 20 \Msun\ stars. Both SN's shock breakout reach their peak luminosity within only one hour. In Figure \ref{fig17:shock breakout}-(a) ($13\;\Msun\;/\;\Omega\;=\;0.2\;/\;Z\;=\;0.0002\;/\;0.6\;\foe\;/\;M_{\rm Ni}\;=\;0.05\;\Msun$), it possesses the most luminous status with one-hour duration. On the other hand, Figure \ref{fig17:shock breakout}-(b) ($20\;\Msun\;/\;\Omega\;=\;0.2\;/\;Z=\;0.02\;/\;0.6\;\foe\;/\;M_{\rm Ni}\;=\;0.02\;\Msun$) has a similar peak luminosity but with a longer duration that make the peak look broader. Features of breakout LC presented here are affected by stellar mass and metallicity in the atmosphere. 

\section{Light Curves from Model and Observation}

\subsection{Light Curves}

After the shock breakout, radiation gradually diffuses from the expanding RSG envelope, producing a long-lasting emission \citep{Nakar&Sari(2010)}. When the star explodes, the optical depth of ejecta decreases as it expands and remains opaque on a day timescale. The SN's luminosity increases as its photosphere expands until it enters a plateau phase at about 100 - 120 days. 

After about 100 days, the luminosity of the SN drops exponentially. During this decaying phase, the primary energy source of radiation comes from the \Co\ decay that originates from \Ni\ formed during the explosion \citep{Falk&Arnett(1973), Srinivasan(2014)}. The energy released from the \Co\ decays through emitting high-energy gamma rays. While the gamma rays scatter within the dense ejecta, it finally heats the ejecta to $5,000 - 10,000\;\K$. Therefore, the spectra of SN emission fall primarily in the visible and infrared wavelengths, and the energy released in the radioactive decay of \Co\ to \Fe\ becomes the primary power source of luminescence of Type II SNe \citep{Falk&Arnett(1977), Chevalier&Fransson(1994), Srinivasan(2014)}. In the next sections, we present comprehensive LC models and compare them with the observations. We use a quantitative assessment of the agreement between simulation and observation, enhancing understanding of the progenitor stars of supernovae.

\subsection{Method of Fitting Observed Light Curves with Models}
We fit the observational data with our grid of LC models based on the two separating periods by dividing the LC into two phases: the plateau phase is the duration from the emergence of LCs till when LC becomes linear decay, and the decay phase starts from the linear decay of LC.

\begin{enumerate}
    
\item Plateau-phase fitting \\
The formation of the LC in the early phase (about 100 - 120 days after the explosion) mainly comes from the energy release of the SN explosion, which shows a plateau shape. Additionally, mass loss affected by initial mass and rotation may also impact the duration of the plateau; thus, our models of various initial masses, rotational rates ($\Omega$), and explosion energies are used to fit with observation and then search for the best-fit model.

\item Decay-phase fitting\newline
The luminosity of linear decay in LCs mainly comes from the \Co\ decay. So, we follow up the best-fit models from plateau-phase fitting and use their decay-phase LCs to obtain the best-fit nickel mass ($M_{\rm Ni}$).

\end{enumerate}

\begin{figure*}[t]
    \centering
    \subfigure[]{
        \includegraphics[scale=0.35]{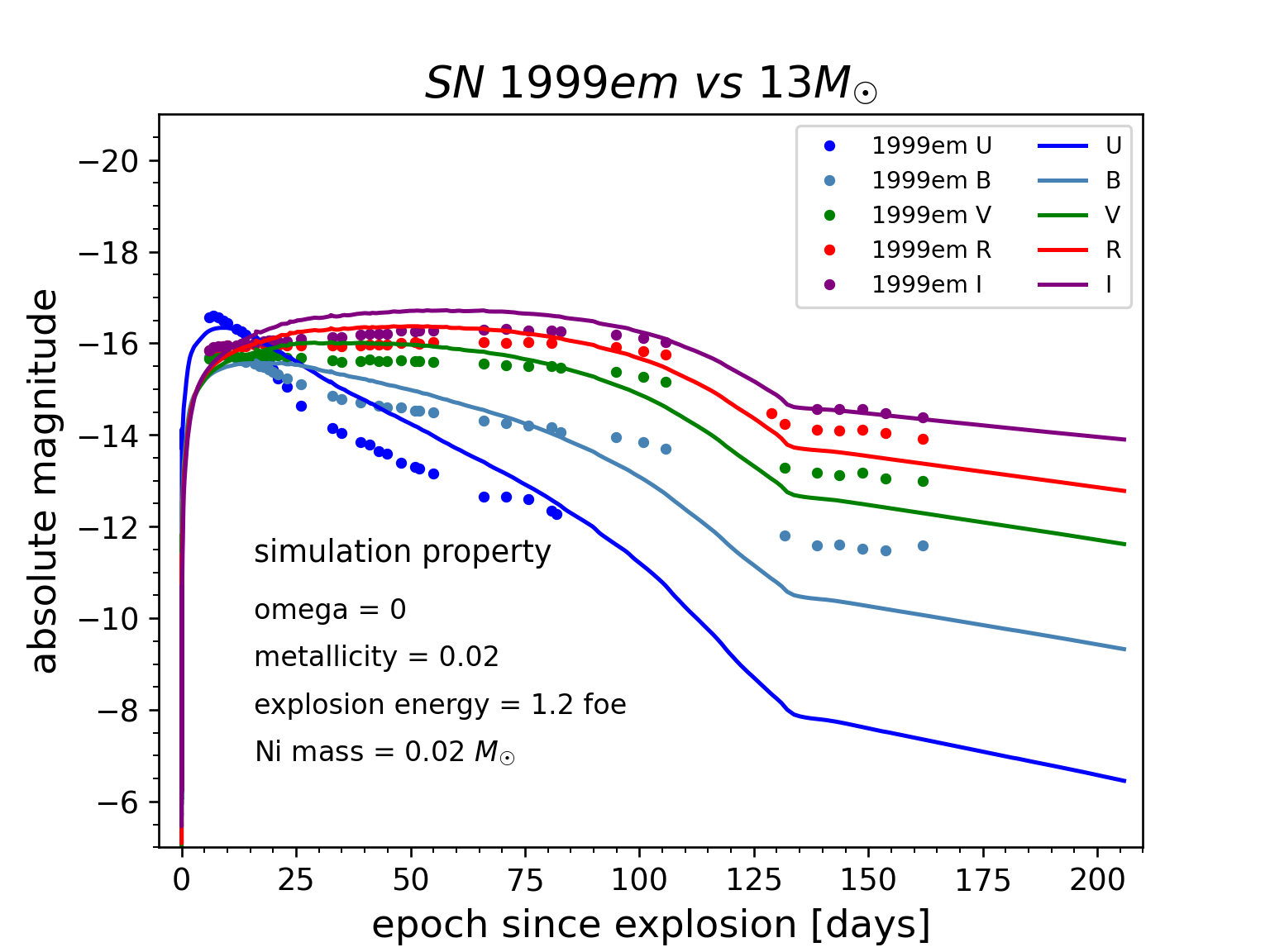}
        }
    \subfigure[]{
        \includegraphics[scale=0.35]{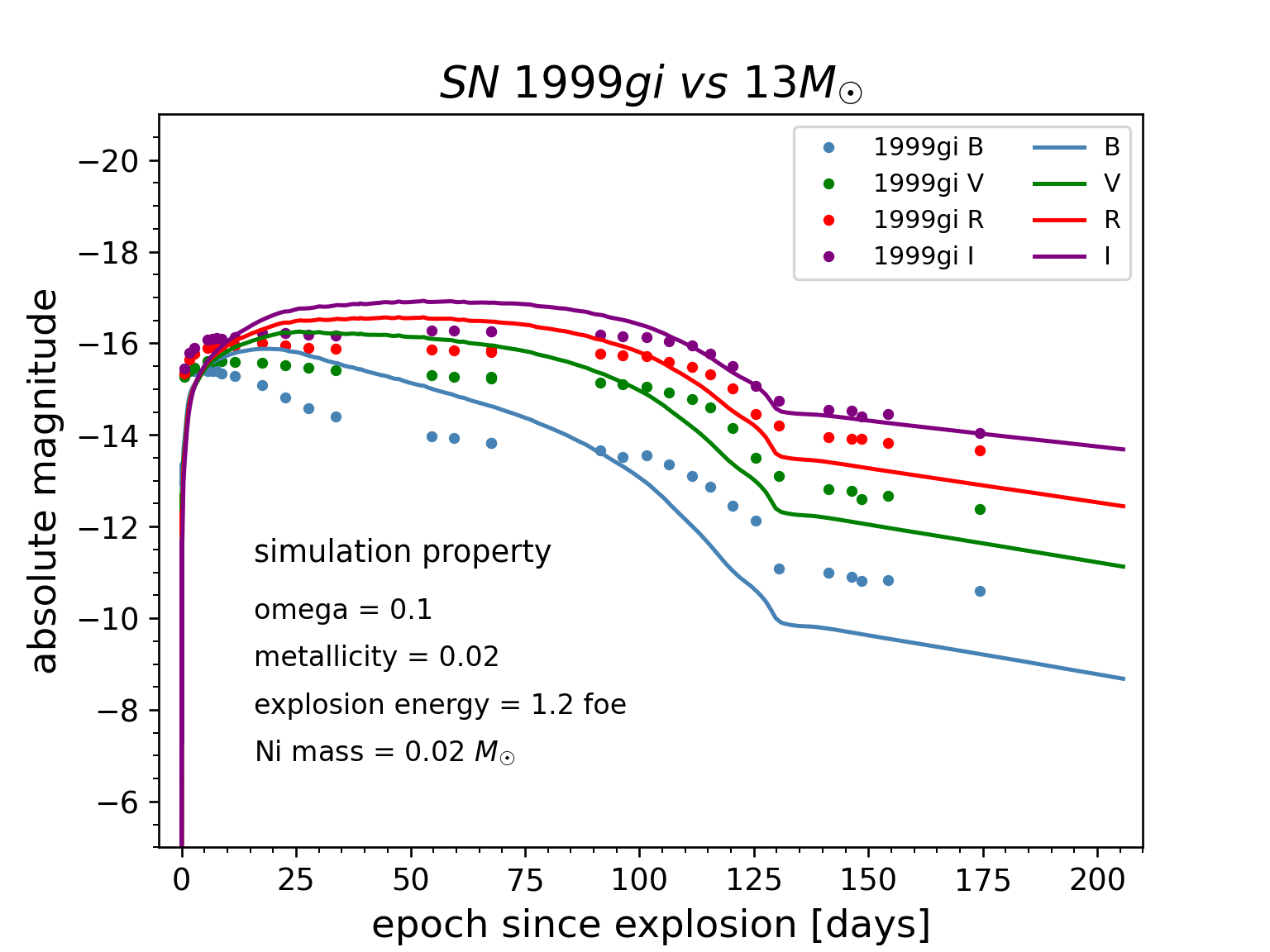}
        }
    \subfigure[]{
        \includegraphics[scale=0.35]{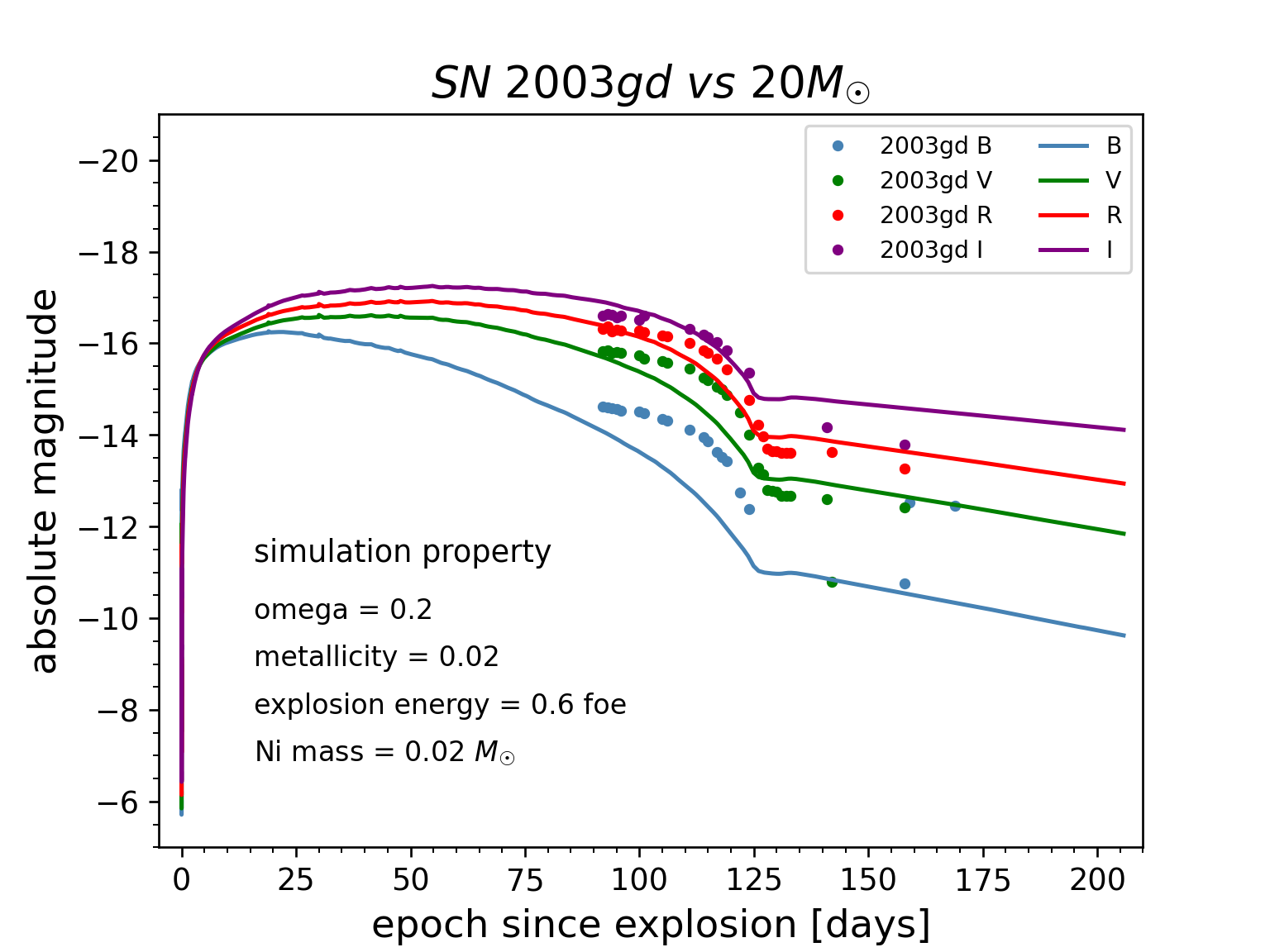}
        }
    \subfigure[]{
        \includegraphics[scale=0.35]{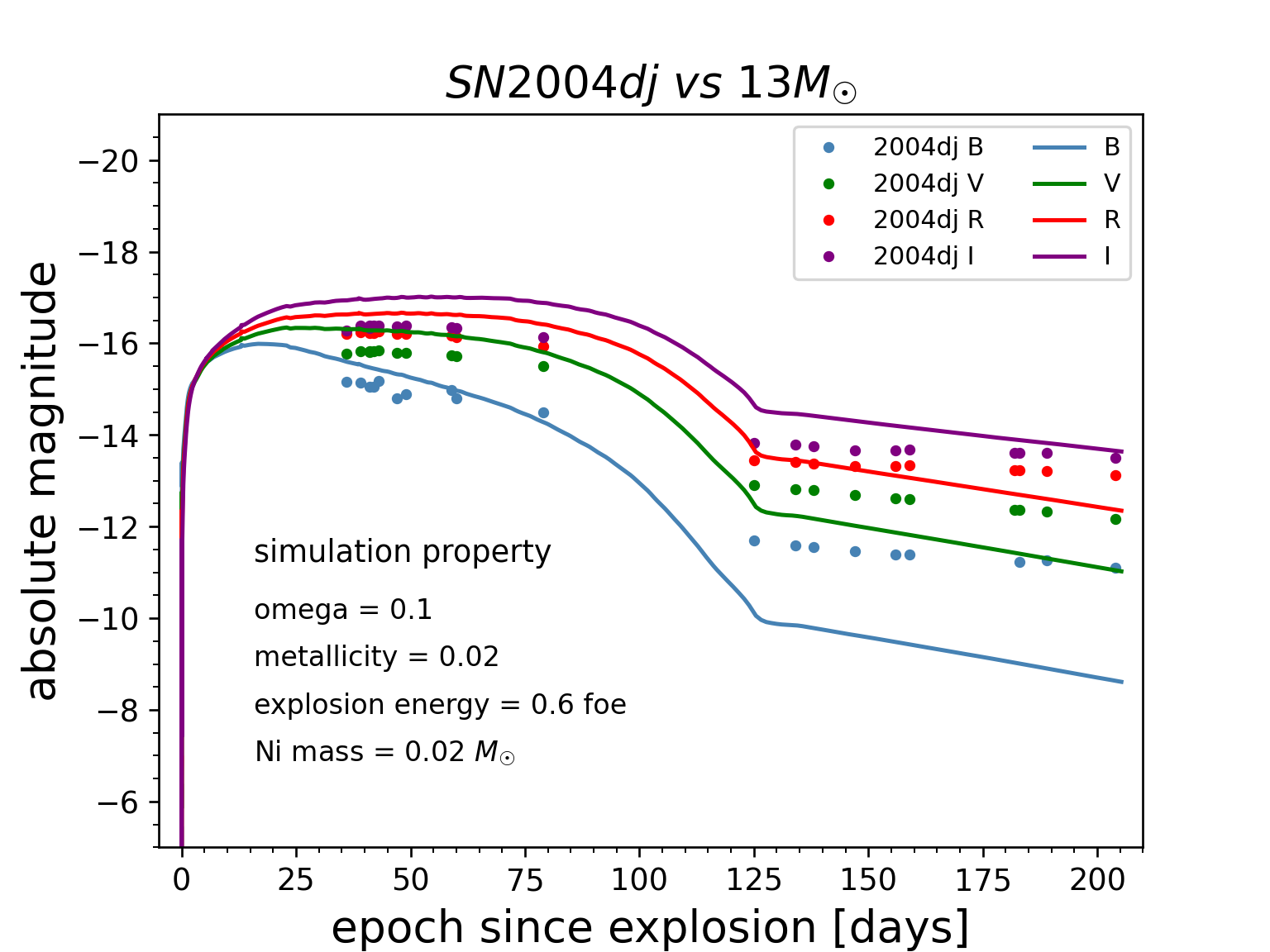}
        }
    \subfigure[]{
        \includegraphics[scale=0.35]{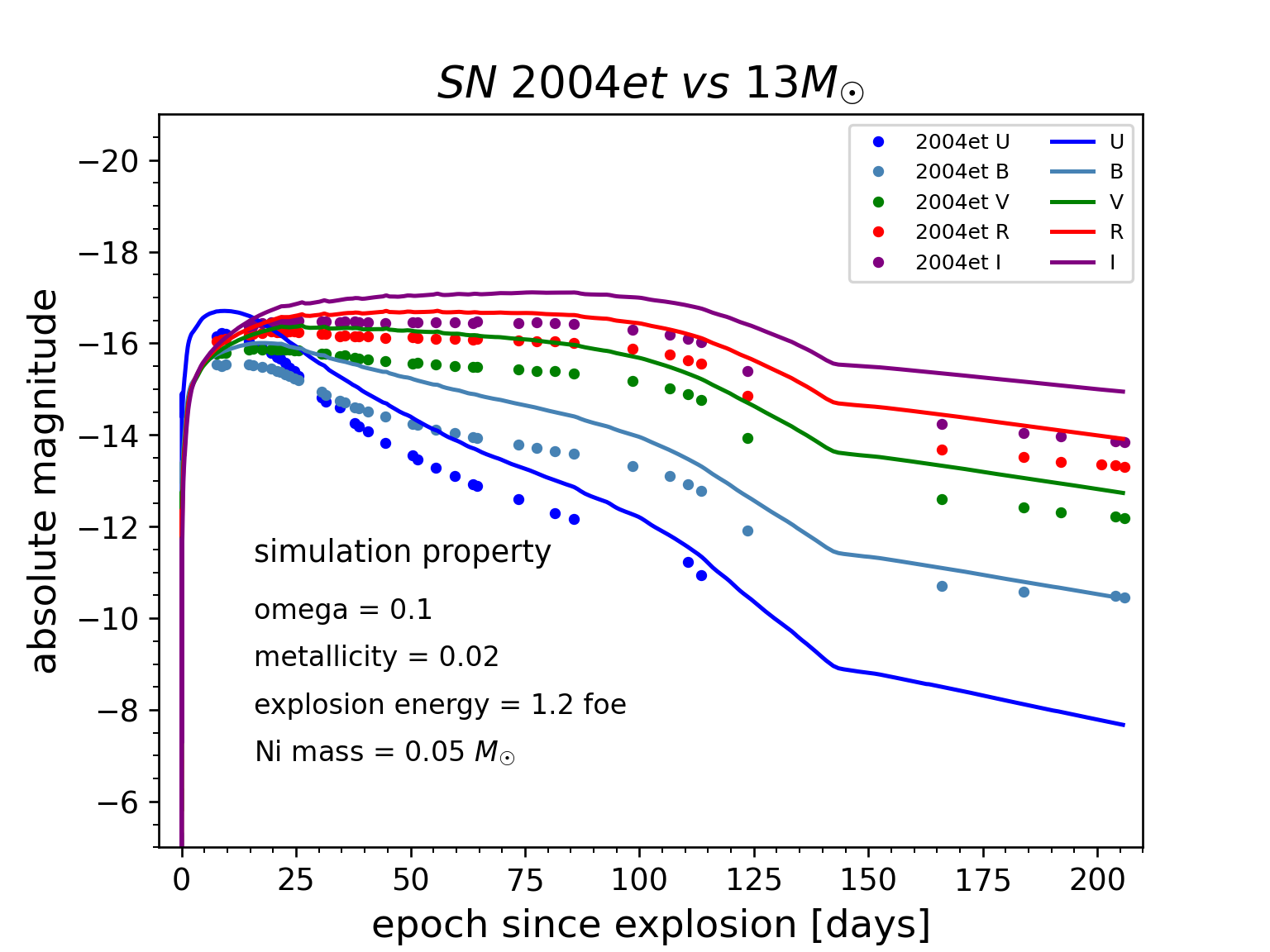}
        }
    \subfigure[]{
        \includegraphics[scale=0.35]{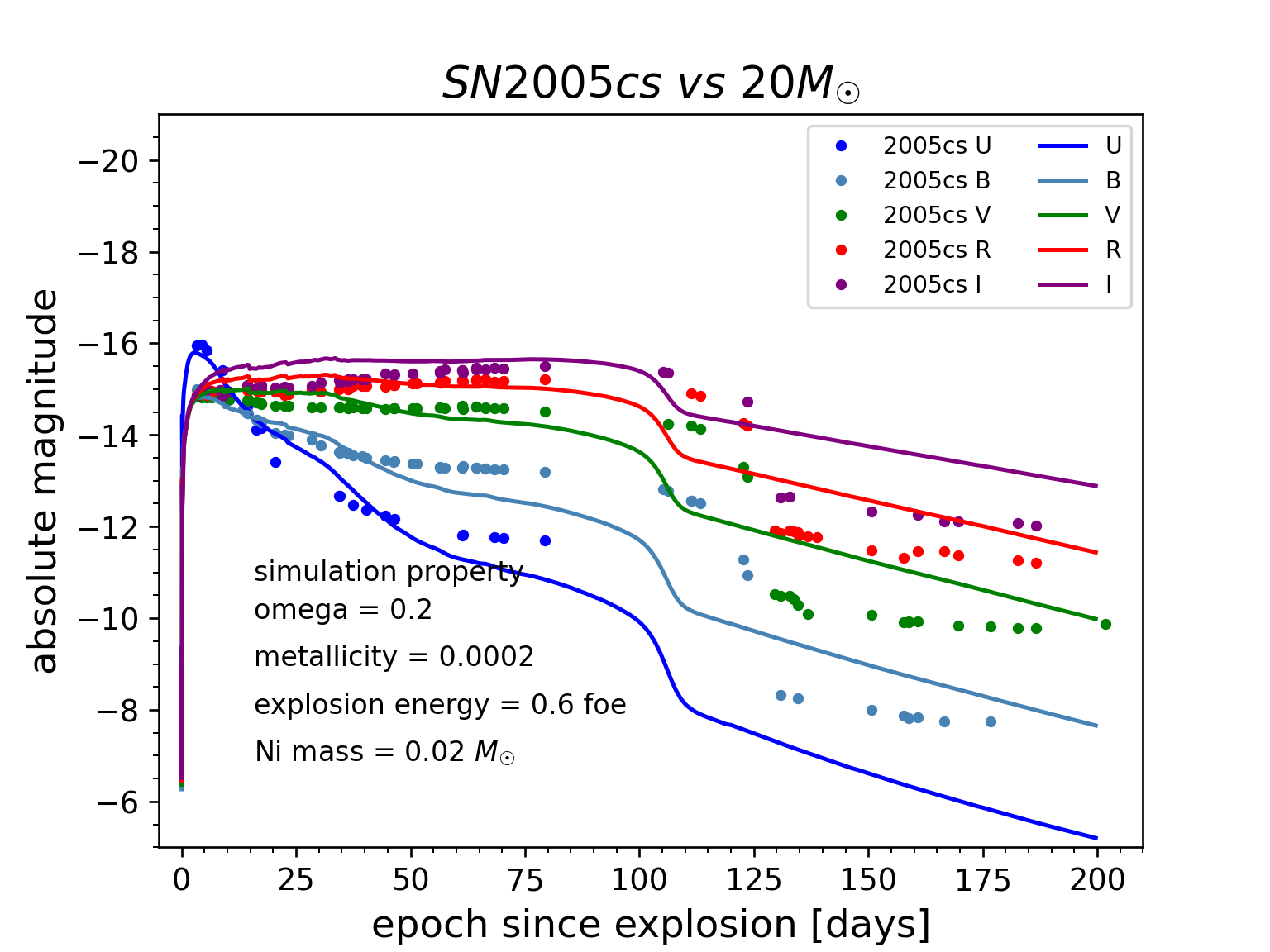}
        }
    \subfigure[]{
        \includegraphics[scale=0.35]{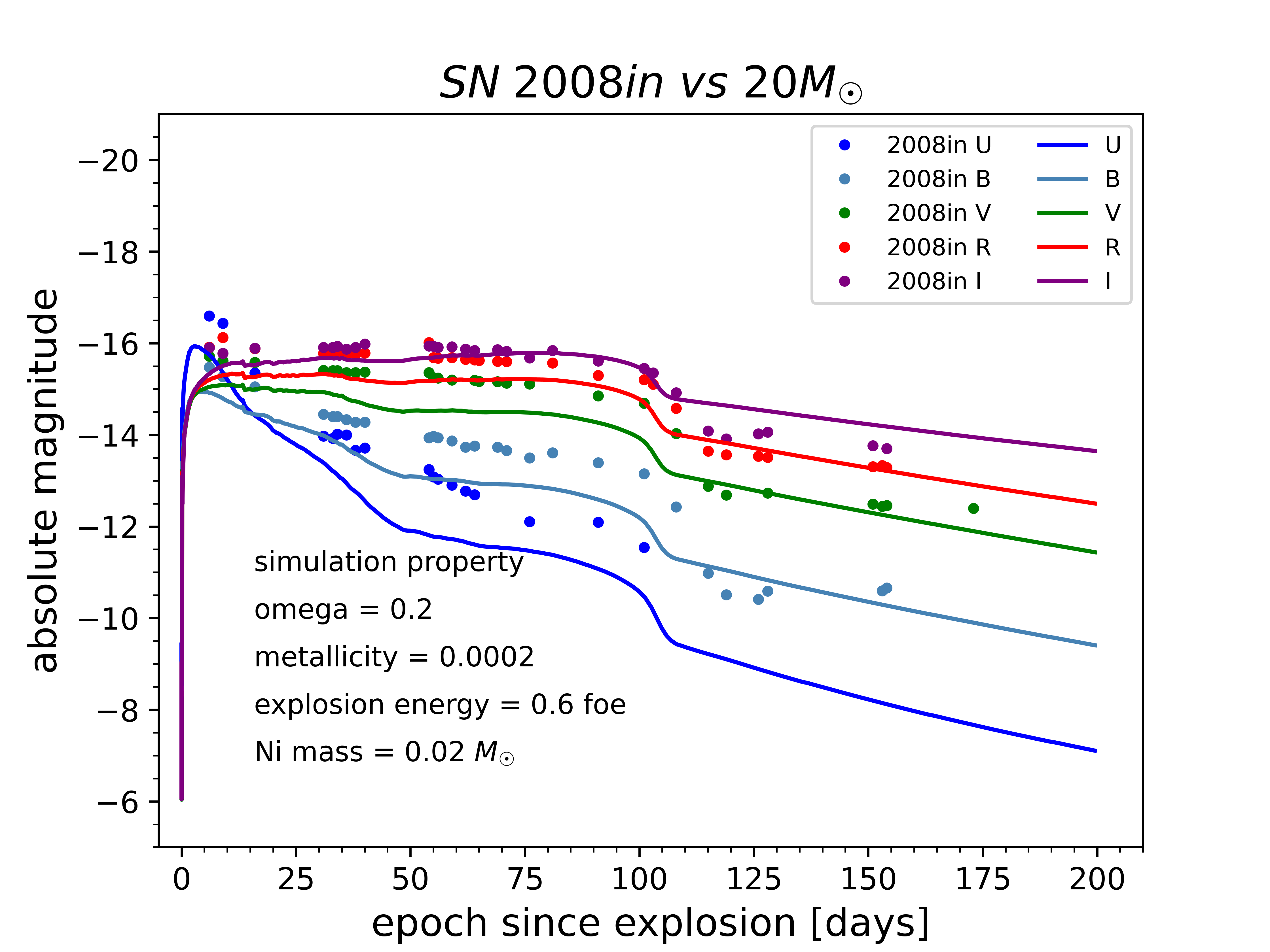}
        }
    \subfigure[]{
        \includegraphics[scale=0.35]{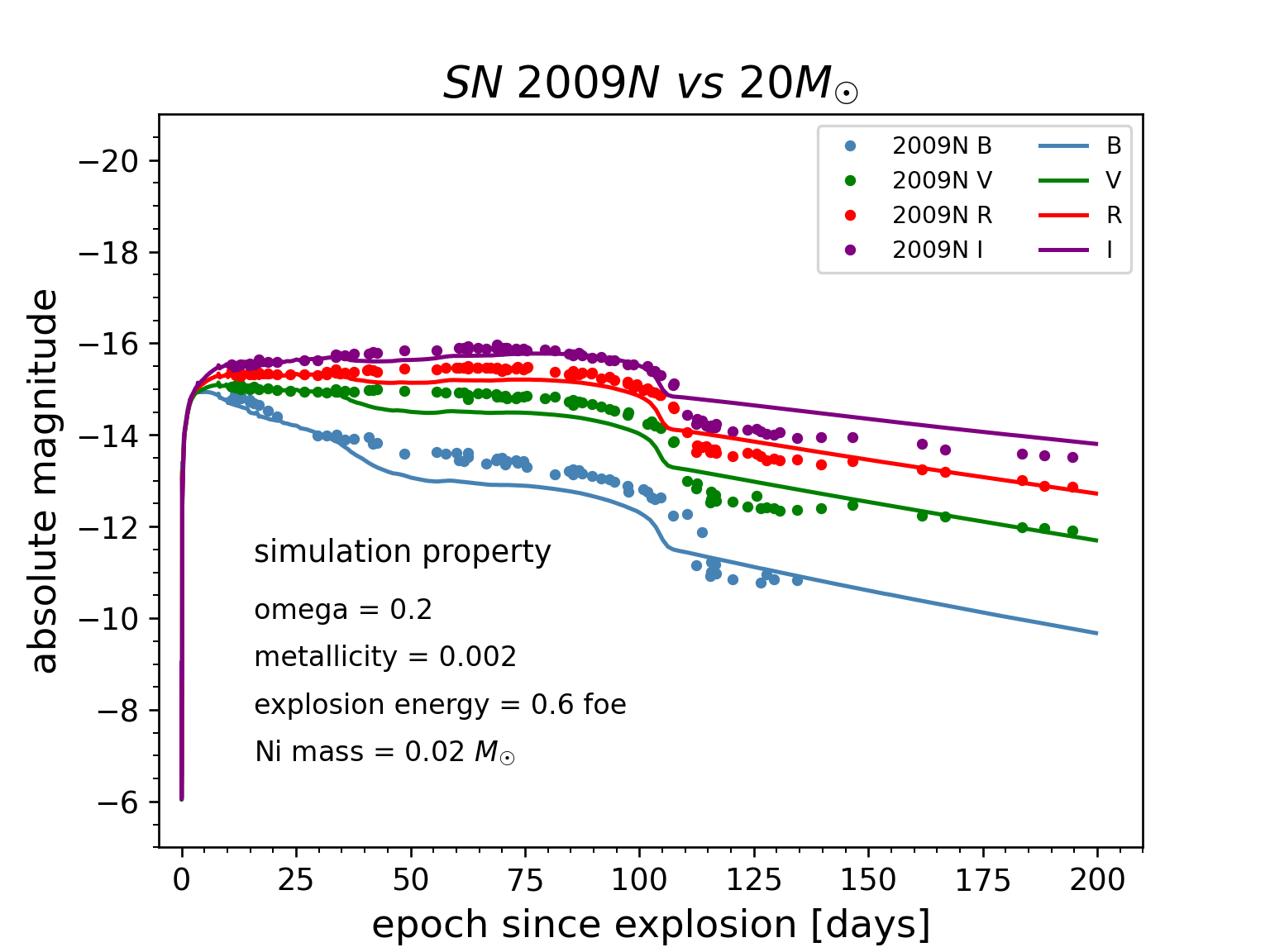}
        }
    \subfigure[]{
        \includegraphics[scale=0.35]{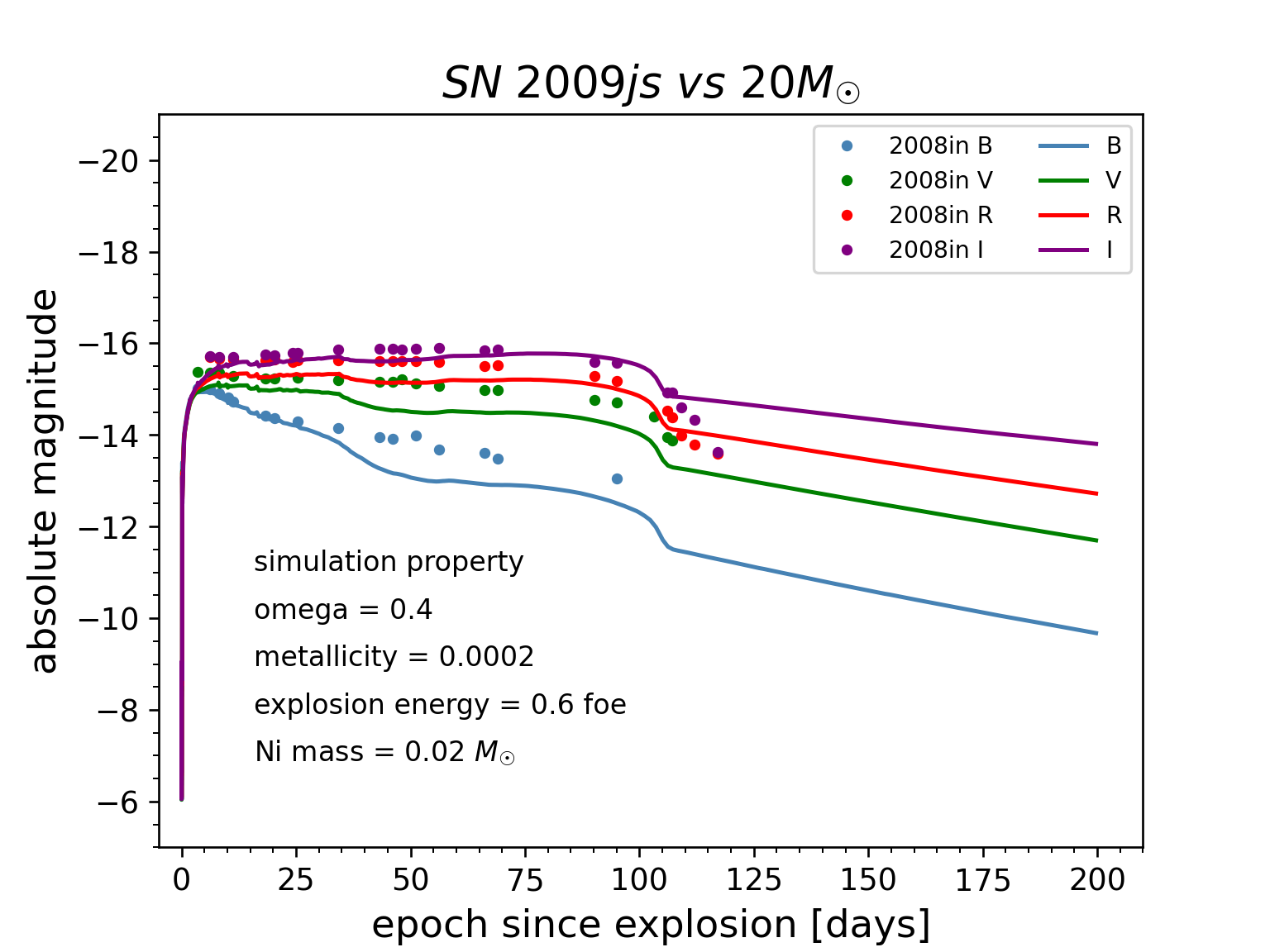}
        }
    \subfigure[]{
        \includegraphics[scale=0.35]{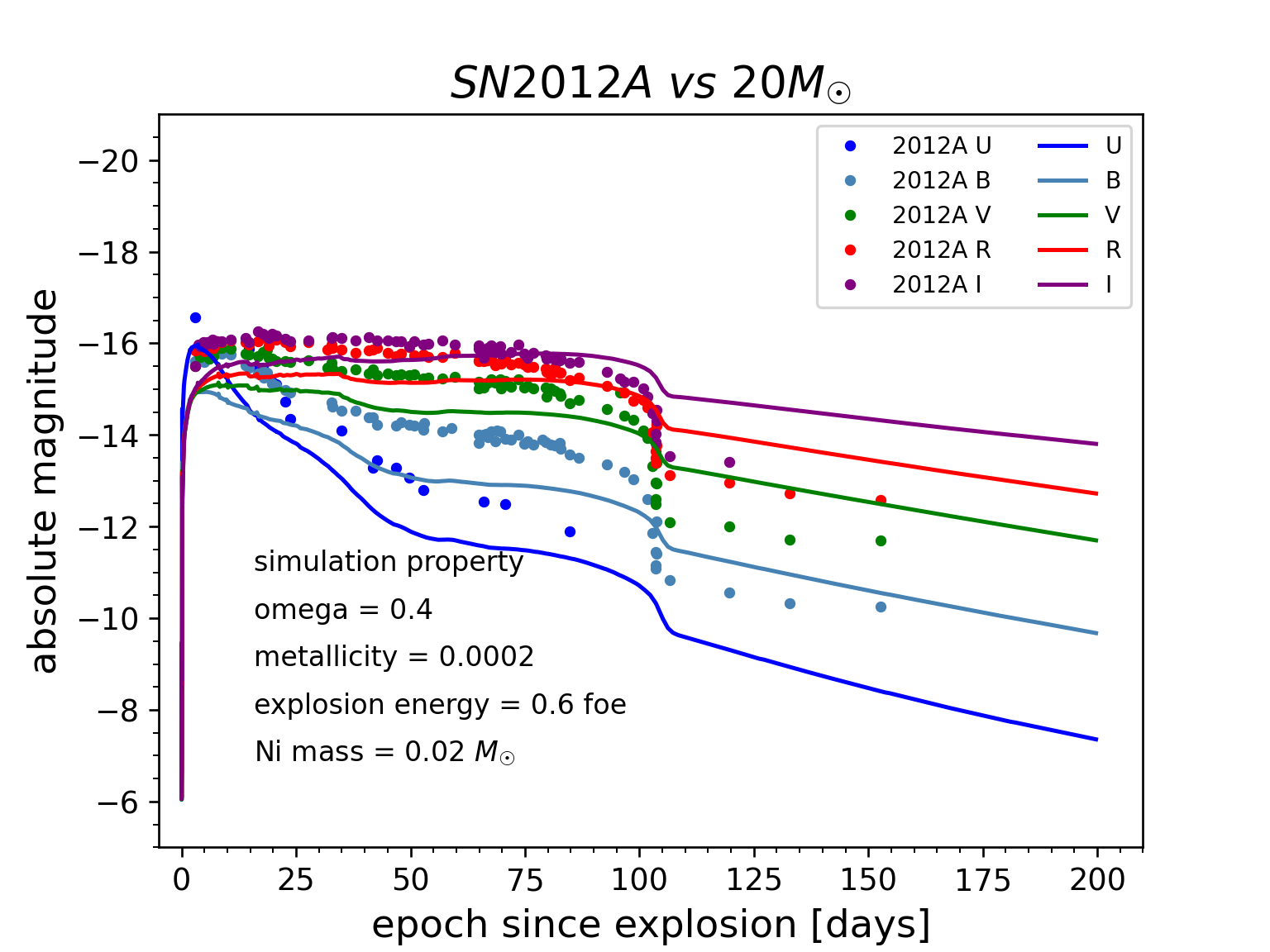}
        }
    \subfigure[]{
        \includegraphics[scale=0.35]{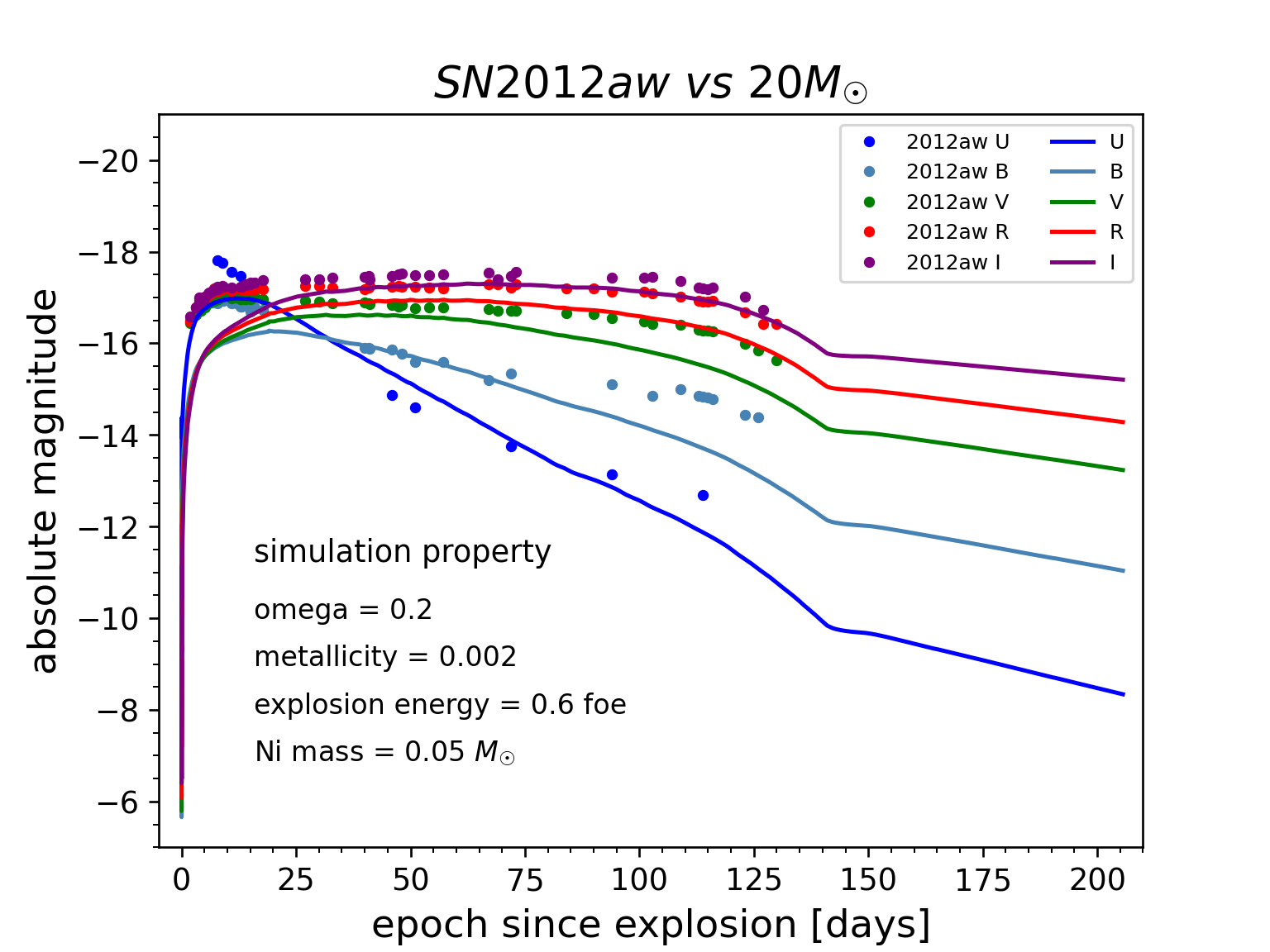}
        }
    \subfigure[]{
        \includegraphics[scale=0.35]{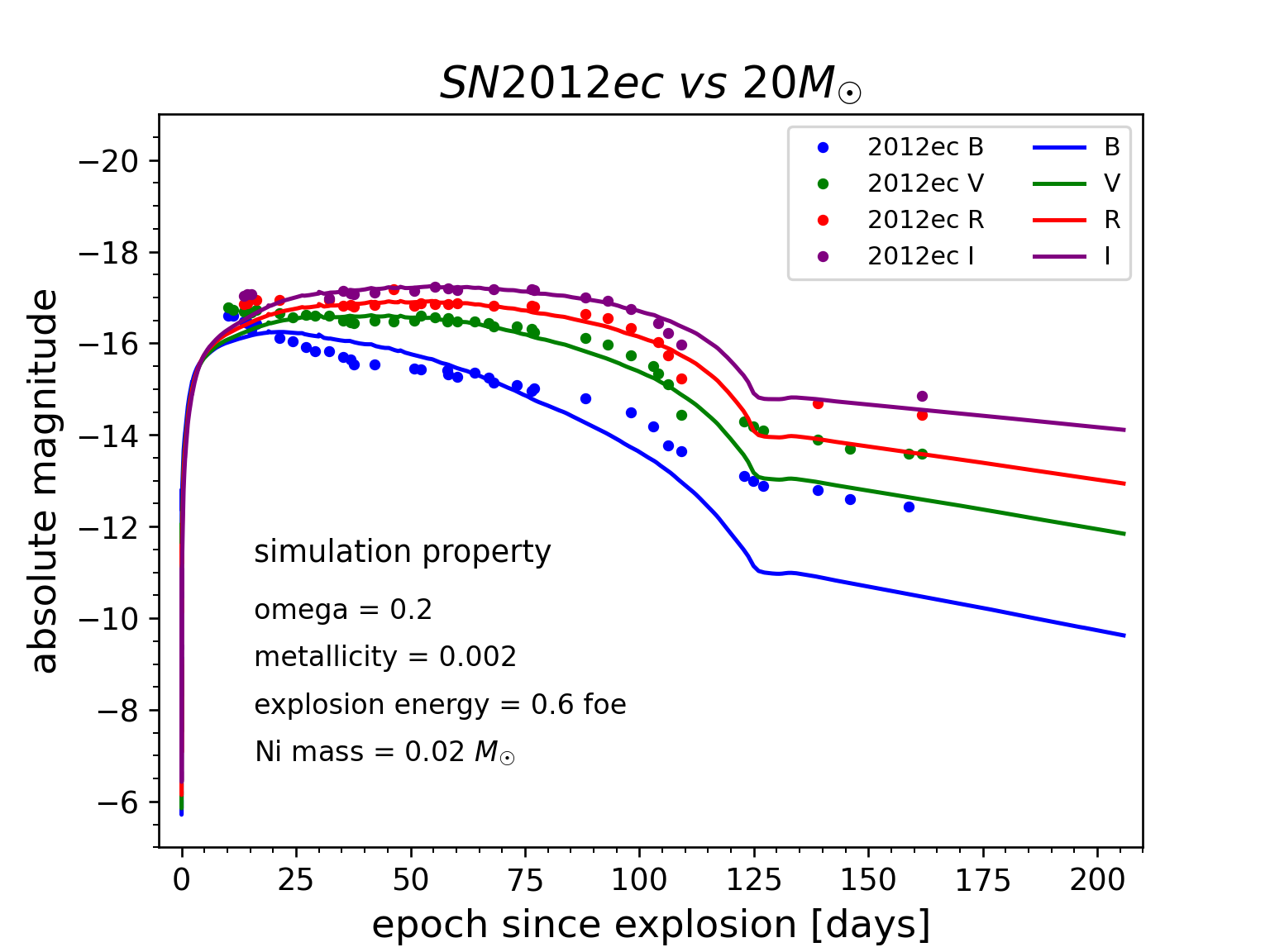}
        }
    \caption{The LC comparison between models and observation within the two hundred days after the explosion.}
    \label{fig15:light curve}
\end{figure*}

The least squares method is employed to reconcile our simulated LC of the SN with the observational data effectively. This approach minimizes the sum of the squared differences between the simulation's predicted values and the observed data points. To find the optimal parameter space that best aligns with the observational data, we iteratively adjust the parameters of our simulation model, such as progenitor's mass, rotation velocity, metallicity, explosion energy, and $M_{\rm Ni}$ (see Table \ref{tab5:simulation result} and Figure \ref{fig15:light curve}). Through this meticulous fitting process, we can quantitatively assess the agreement between our simulation and observational data, thus enhancing our understanding of the SN's progenitor stars. The error bar of observational data is very small compared to the measured values and illegible in our plots, so we choose not to show the error bars in the fitting plot of Figure \ref{fig15:light curve}.

\begin{table}[]
\hskip-2cm
\resizebox{20.2cm}{!}{%
\begin{tabular}{|lccccc|cccccl|}
\hline
\multicolumn{6}{|c|}{Our best fitting value}                                                                     & \multicolumn{6}{c|}{Reference best fitting value}                                                                                                  \\ \hline
\multicolumn{1}{|l|}{Progenitor star}         & Initial Mass  & $\Omega$ & Metallicity & Explosion Energy & $M_{\rm Ni}$         & Initial Mass                 & $\Omega$ & Metallicity & Explosion Energy            & $M_{\rm Ni}$                      & Reference                     \\ \hline
\multicolumn{1}{|l|}{$SN\;1999em$} & $13\;\Msun$ & 0.0      & 0.02        & $1.2\;\foe$       & $0.02\;\Msun$ & -                            & -        & -           & -                           & -                            & \cite{Leonardetal.(2002)}    \\
\multicolumn{1}{|l|}{$SN\;1999gi$} & $13\;\Msun$ & 0.1      & 0.02        & $1.2\;\foe$       & $0.02\;\Msun$ & $15_{-3}^{+5}\;\Msun$      & -        & -           & -                           & -                            & \cite{Leonardetal.(2002)-1}  \\
\multicolumn{1}{|l|}{$SN\;2003gd$} & $20\;\Msun$ & 0.2      & 0.02        & $0.6\;\foe$       & $0.02\;\Msun$ & $8_{-2}^{+4}\;\Msun$       & -        & -           & -                           & $0.016\;\Msun$             & \cite{Hendryetal.(2005)}     \\
\multicolumn{1}{|l|}{$SN\;2004A$}  & $13\;\Msun$ & 0.2      & 0.0002      & $0.6\;\foe$       & $0.05\;\Msun$ & -                            & -        & -           & -                           & -                            & \cite{Tsvetkov(2008)}        \\
\multicolumn{1}{|l|}{$SN\;2004dj$} & $13\;\Msun$ & 0.1      & 0.02        & $0.6\;\foe$       & $0.02\;\Msun$ & $>20\;\Msun$               & -        & -           & $0.86_{-0.49}^{+0.89}\;\foe$ & $0.02\;\pm\;0.01\;\Msun$   & \cite{Vinketal.(2006)}       \\
\multicolumn{1}{|l|}{$SN\;2004et$} & $13\;\Msun$ & 0.1      & 0.02        & $1.2\;\foe$       & $0.05\;\Msun$ & $20\;\Msun$                & -        & -           & $1.20_{-0.30}^{+0.38}\;\foe$ & $0.06\;\pm\;0.02\;\Msun$   & \cite{Sahuetal.(2006)}       \\
\multicolumn{1}{|l|}{$SN\;2005cs$} & $20\;\Msun$ & 0.2      & 0.0002      & $0.6\;\foe$       & $0.02\;\Msun$ & $8\;-\;15\;\Msun$          & -        & -           & $0.3\;\foe$                  & $0.003\;\Msun$             & \cite{Pastorelloetal.(2009)} \\
\multicolumn{1}{|l|}{$SN\;2007od$} & - & -      & -        & -       & - & $9.7\;-\;11\;\Msun$        & -        & -           & $0.5\;\foe$                  & $0.02\;\Msun$              & \cite{Inserraetal.(2011)}    \\
\multicolumn{1}{|l|}{$SN\;2008in$} & $20\;\Msun$ & 0.2      & 0.0002      & $0.6\;\foe$       & $0.02\;\Msun$ & $<20\;\Msun$               & -        & -           & $0.54\;\foe$                 & $0.015\;\Msun$             & \cite{Royetal.(2011)}        \\
\multicolumn{1}{|l|}{$SN\;2009bw$} & - & -      & -        & -       & - & $11\;-\;15\;\Msun$          & -        & -           & $0.3\;\foe$                  & $0.022\;\Msun$             & \cite{Inserraetal.(2012)}    \\
\multicolumn{1}{|l|}{$SN\;2009js$} & $20\;\Msun$ & 0.4      & 0.0002      & $0.6\;\foe$       & $0.02\;\Msun$ & $8.5_{-1.5}^{+6.5}\;\Msun$ & -        & -           & -                           & $0.054\;\pm\;0.013\;\Msun$ & \cite{Gandhietal.(2013)}     \\
\multicolumn{1}{|l|}{$SN\;2009N$}  & $20\;\Msun$ & 0.4      & 0.0002      & $0.6\;\foe$       & $0.02\;\Msun$ & $13\;-\;13.5\;\Msun$       & -        & -           & $0.48\;\foe$                 & $0.020\;\pm\;0.004\;\Msun$ & \cite{Taktsetal.(2014)}      \\
\multicolumn{1}{|l|}{$SN\;2012A$}  & $20\;\Msun$ & 0.2      & 0.0002      & $0.6\;\foe$       & $0.02\;\Msun$ & $10.5_{-2}^{+4.5}\;\Msun$  & -        & -           & $0.48\;\foe$                 & $0.011\;\pm\;0.004\;\Msun$ & \cite{Tomasellaetal.(2013)}  \\
\multicolumn{1}{|l|}{$SN\;2012aw$} & $20\;\Msun$ & 0.2      & 0.002       & $0.6\;\foe$       & $0.05\;\Msun$ & -                            & -        & -           & $1.5\;\foe$                  & $0.056\;\pm\;0.013\;\Msun$ & \cite{Dall'Oraetal.(2014)}   \\
\multicolumn{1}{|l|}{$SN\;2012ec$} & $20\;\Msun$ & 0.2      & 0.002       & $0.6\;\foe$       & $0.02\;\Msun$ & $14\;-\;22\;\Msun$         & -        & -           & $1.5\;\foe$                  & $0.040\;\pm\;0.015\;\Msun$ & \cite{Barbarinoetal.(2015)}  \\
\multicolumn{1}{|l|}{$SN\;2013ab$} & $15\;\Msun$ & 0.4      & 0.02        & $0.6\;\foe$       & $0.05\;\Msun$ & $9\;\Msun$                 & -        & -           & $0.35\;\foe$                 & $0.064\;\Msun$             & \cite{Boseetal.(2015)}       \\
\multicolumn{1}{|l|}{$SN\;2013ej$} & $20\;\Msun$ & 0.4      & 0.02        & $1.2\;\foe$       & $0.05\;\Msun$ & $12\;-\;13\;\Msun$         & -        & -           & $0.7\;\foe$                  & $0.02\;\pm\;0.01\;\Msun$   & \cite{Huangetal.(2015)}      \\ \hline
\end{tabular}%
}
\caption{The best parameters for observed SNe based on our models. The best-fit parameters for $SN\;2007od$ and $SN\;2009bw$ are blank because our models cannot fit well.}
\label{tab5:simulation result}
\end{table}

\subsection{Fitting Results}

Compared with the observation, the fitting results in the plateau-phase are roughly reasonable for SNe IIP, in which linear decay occurs 100 - 120 days after the explosion. However, in $SN\;1999em$, $SN\;1999gi$, $SN\;2004dj$, $SN\;2004et$ (Figures \ref{fig15:light curve}-(a), \ref{fig15:light curve}-(b), \ref{fig15:light curve}-(d), \ref{fig15:light curve}-(e)), we find that the $M$ of our models in plateau-phase has discrepancy from the observation, the results show that our models are brighter than those. The trend of the observed brightness approximates a straight horizontal line, which means the brightness is always the same in the plateau-phase. On the other hand, our model of $SN\;2008in$, $SN\;2009js$, $SN\;2012A$ is fainter than the observations, which also have the same brightness in the plateau-phase.

Since the plateau region in the LC of a SN IIP comes from the hydrogen recombination of the SN's outer layers after the explosion, when the shock wave generated by the explosion propagates outward through the SN, it heats the outer layer to more than $10^{5}$K. It ionizes all hydrogen, contributing a significant opacity to a photosphere \citep{Goldberg(2022)}. When the ionized hydrogen in the outer layer cools down to the recombination temperature of $5,500\;-\;6,000$K and recombines back to neutral hydrogen \citep{Dessartetal.(2013), Goldberg(2022)}, the photons emitted from the recombined area inside the envelope can escape. 

The LC forms a plateau shape since this temperature remains roughly constant after the photosphere passes through the hydrogen envelope. Based on Table \ref{tab6:metallicity}, we discover that those own hydrogen with higher ratio (i.e., less metallicity $Z$) form a lighter LC than the observation in the same explosion energy (see Figures \ref{fig15:light curve}-(a), \ref{fig15:light curve}-(b)). This result suggests that the hydrogen mass fraction can profoundly affect the SN's brightness, and metallicity plays a key role in the mass loss of the hydrogen envelope of a star \citep{Ouetal.(2023)}. The hydrogen ratio and metallicity should be adjusted more specifically.

According to the decay-phase results, the LCs of $20\;\Msun$ with different explosion energies will decay exponentially in 100 - 120 days. However, $SN\;2012aw$ show no signs of decay until 130 days (Figure \ref{fig15:light curve}-(k)). We speculate that the explosion energy of the SN is relatively weak, and the ratio of hydrogen on the surface is higher than expected (see Table \ref{tab6:metallicity}). Thus, the radiation emitted from recombination can last longer than usual SN, which explains the long duration in the plateau, and the hydrogen envelope and metallicity determine the turning point between plateau and decay.

Photospheric velocity ($v_{\rm ph}$) is a critical parameter that sheds light on the dynamics and energetics of SN by characterizing the speed at which this outermost material layer moves away from the central remnant of the exploded star. \STELLA\ finds the photosphere by calculating the location where optical depth $\tau\;=\;\frac{2}{3}$ using the Rosseland mean opacity (grey = overall frequency groups) and the photospheric velocity is defined as the fluid velocity at that location in the ejecta e.g., \cite{Kasen&Woosley(2009), Goldbergetal.(2019)}. Meanwhile, the observed $v_{\rm ph}$ comes from several emission lines of hydrogen and iron, e.g., \cite{Boseetal.(2015)}, and $v_{\rm ph}$ from different lines varies much. We compare the averaged $v_{\rm ph}$ from various emission lines with that from \STELLA.

From Table \ref{tab6:metallicity}, the overall theoretical \(v_{\rm ph}\) of our best-fit models is smaller than the observed values (Table \ref{tab1:CCSN info}) by 1500–2500 \(\text{km}\,\text{s}^{-1}\). These discrepancies in \(v_{\rm ph}\) can arise due to various factors related to the complexity of astrophysical phenomena and limitations in modeling techniques. For example, when the shock breaks out on the stellar surface, the influx of radiation can exert pressure and accelerate the ejecta around the photosphere, leading to high \(v_{\rm ph}\). We suspect that \STELLA\ may not properly evolve the dynamics of the radiative shock due to its limited spatial resolution. Furthermore, strong fluid instabilities emerging from the explosion and shock breakout can lead to turbulent mixing of the ejecta, thereby altering \(v_{\rm ph}\) \citep{Chenetal.(2024)}. Unfortunately, modeling such multidimensional mixing is beyond the capability of \STELLA.

\begin{table}[]
\hskip-1cm
\centering
\resizebox{9cm}{!}{%
\begin{tabular}{|l|ccc|}
\hline
Progenitor star         & Hydrogen mass fraction X & Metallicity Z  & $v_{\rm ph}$ (\kms)  \\ \hline
$SN\;1999em$ & 0.6654441     & 0.0201196     &1916  \\
$SN\;1999gi$ & 0.6654441     & 0.0201196     &2044  \\
$SN\;2003gd$ & 0.0200412     & 0.0200412     &2637  \\
$SN\;2004A$  & 0.6725015     & 0.0002018     &2254  \\
$SN\;2004dj$ & 0.6654441     & 0.0201196     &2358  \\
$SN\;2004et$ & 0.6654441     & 0.0201196     &2076  \\
$SN\;2005cs$ & 0.6940123     & 0.0002022     &1231  \\
$SN\;2007od$ & -     & -     &- \\
$SN\;2008in$ & 0.6940123     & 0.0002022     &1231  \\
$SN\;2009bw$ & -     & -     &- \\
$SN\;2009js$ & 0.6940123     & 0.0002022     &1383  \\
$SN\;2009N$  & 0.6940123     & 0.0002022     &1383  \\
$SN\;2012A$  & 0.6940123     & 0.0002022     &1231  \\
$SN\;2012aw$ & 0.5051498     & 0.0019692     &2296  \\
$SN\;2012ec$ & 0.5051498     & 0.0019692     &2850  \\
$SN\;2013ab$ & 0.6249238     & 0.0201112     &3080  \\
$SN\;2013ej$ & 0.5295910     & 0.0199381     &4115  \\ \hline
\end{tabular}%
}
\caption{Physical parameters of our best-fit models right before the explosion.}

\label{tab6:metallicity}
\end{table}

\subsection{Analysis of Color Index}

\begin{figure*}[ht]
    \centering
    	\includegraphics[scale=0.7]{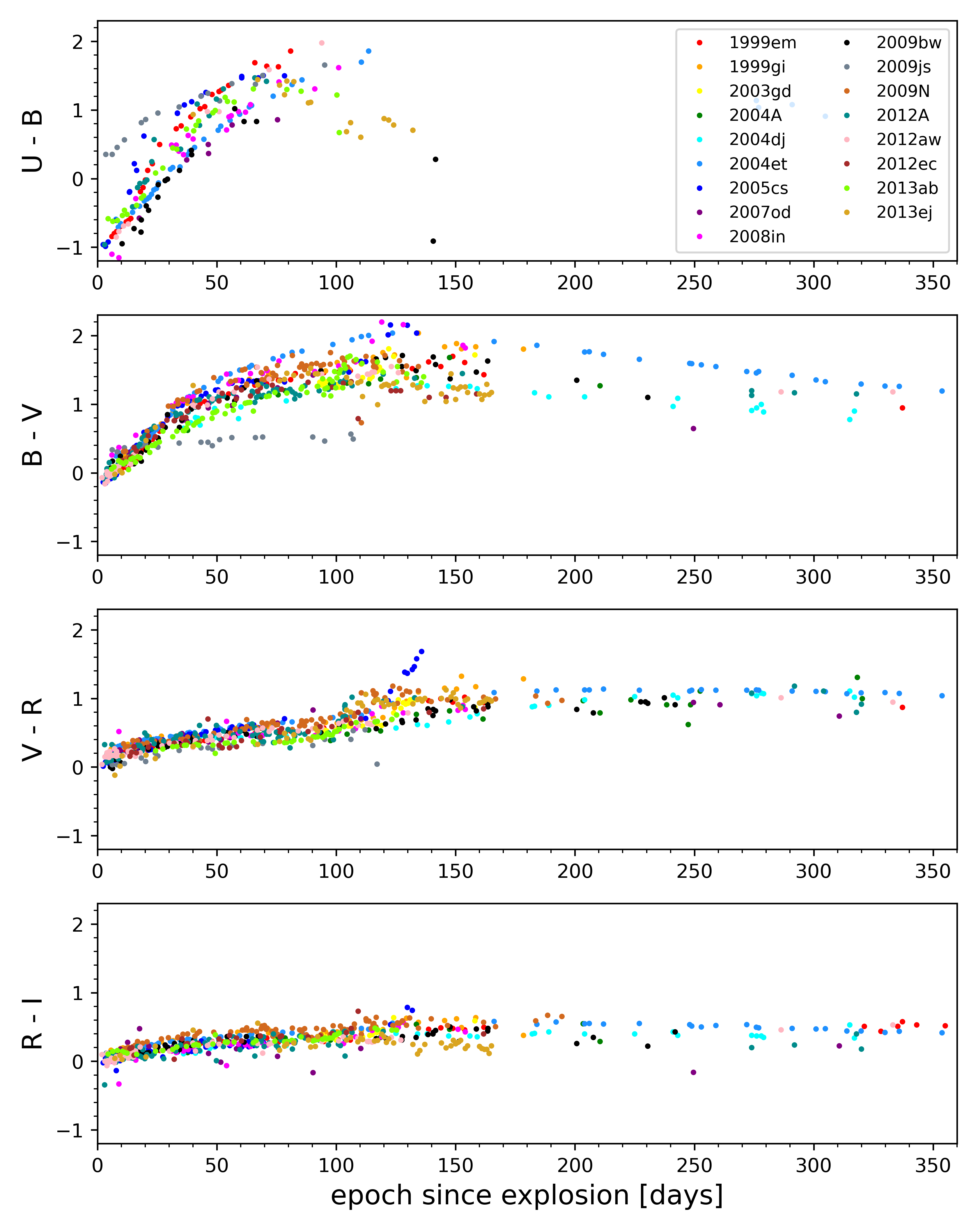}
    \caption{The change of color index of observed SNe IIP in 360 days from history observation.}
    \label{fig14:colorindex}
\end{figure*} 

Figure \ref{fig14:colorindex} shows various SNe IIP color indexes, and their colors have been corrected for the extinction effect of the Milky Way and their host galaxies. These data points show similar features across many stellar properties, and the points are associated with luminosity, expansion velocity, temperature, and $M_{\rm Ni}$, allowing the expansion and cooling behavior of the outer shell to be studied based on the evolution of color index. Each panel in Figure \ref{fig14:colorindex} shows that features are similar up to 100 days after the explosion except a more rapid change for $U\;-\;B$ color index that suggests the color was blue at first but quickly turned red after 50 days since explosions. 

During the early phase of the explosion, the star's outer layers are expelled rapidly, producing radiation across the electromagnetic spectrum, including $U$ and $B$ bands. This phase contributes to the initial increase in brightness and blueness in $U\;-\;B$ color index \citep{Srinivasan(2014), Waxman&Katz(2017)}. Then, the ejecta expands rapidly, causing SN to cool down. As the temperature goes down, the dominant source of radiation shifts from the ultraviolet to the visible and infrared wavelengths. This shift decreases the $U\;-\;B$ color index as the SN evolves.

As the SN expands and cools further, the ionized hydrogen starts to recombine and forms a photosphere. The photosphere's characteristics, such as its temperature and composition, influence the observed colors of the SN. Changes in the composition and temperature of the photosphere contribute to variations in the color index over time. Furthermore, The composition of the ejecta and blanketing effect caused by absorption lines from various elements in the SN's spectrum also affect the observed colors. As the SN evolves and different elements dominate the spectrum, the $U\;-\;B$ color index may experience fluctuations \citep{Filippenko(1997)}.

In $B\;-\;V$, $V\;-\;R$, and $R\;-\;I$ color index, most of the SNe change less significantly in 100 - 120 days, but in the $B\;-\;V$ color index, we can see that the color index start to decrease, showing a blue trend after 120 days, which is a period when the ejecta is optically thinned and  \Co\ is decaying into \Fe, then SN enters the nebular phase. However, $SN\;2009js$ only slowly increases in the $B\;-\;V$ color index within four months, which suggests that its $T_{\rm eff}$  changes slowly and retains bluish. Furthermore, $SN\;2004et$, $SN\;2005cs$, and $SN\;2008in$ reach the climax in $B\;-\;V$ color index at 110 - 130 days, which shows a sudden transition to reddish color.

The metallicity of progenitor stars can be estimated through photometric and spectroscopic measurements. From the study \cite{Sekiguchi&Fukugita(2000)} and \cite{Ballesteros(2012)}, they have used the modern, least model-dependent ways to derive the relationship between the effective temperature of stars, $B\;-\;V$ index color, metallicity, and surface gravity as below:

\begin{equation}
\begin{split}
\label{eqn11} \log_{10}T_{\rm eff}\;=\;c_{0}\;+\;c_{1}(B-V)_{0}\;+\;c_{2}(B-V)^2_{0}\;+\;c_{3}(B-V)^3_{0}\\
+\;f_{1}[Fe/H]\;+\;f_{2}[Fe/H]^2\;+\;g_{1}\log_{10}g\;+\;h_{1}(B-V)_{0}\log_{10}g\;,
\end{split}
\end{equation}

where $\rm [Fe/H]$ is the metallicity relative to the solar metallicity, \Zsun:

\begin{equation}
\label{eqn12} [Fe/H]\;=\;\log_{10}(\dfrac{N_{Fe}}{N_{H}})_{\ast}\;-\;\log_{10}(\dfrac{N_{Fe}}{N_{H}})_{\odot}\;.
\end{equation}

\begin{table}[ht]
\centering
\resizebox{5.5cm}{!}{%
\begin{tabular}{|c|c|}
\hline
Parameter   & Value                     \\ \hline
$c_{0}$     & $3.939654\;\pm\;0.0115$   \\
$c_{1}$     & $-0.395361\;\pm\;0.0263$  \\
$c_{2}$     & $0.2082113\;\pm\;0.0293$  \\
$c_{3}$     & $-0.0604097\;\pm\;0.0107$ \\
$f_{1}$     & $0.027153\;\pm\;0.00166$  \\
$f_{2}$     & $0.005036\;\pm\;0.000972$ \\
$g_{1}$     & $0.007367\;\pm\;0.00231$  \\
$h_{1}$     & $-0.01069\;\pm\;0.00224$  \\
$log_{10}g$ & 4.3                       \\ \hline
\end{tabular}%
}
\caption{The list of parameter fitting values from Eqn. \ref{eqn11} based on \cite{Sekiguchi&Fukugita(2000)}.}
\label{tab7:parameter}
\end{table}

Based on Table \ref{tab1:CCSN info}, we retrieve the value $(B\;-\;V)$ from observation data and $T_{\rm eff}$ from Eqn. \ref{eqn11} and empirical value from \cite{Pecaut&Mamajek(2013)}, and then to calculate the ratio $\rm [Fe/H]$ when the star is at the end of evolution with parameters shown in Table \ref{tab7:parameter}, which is helpful to us to find the appropriate model to fit (see Table \ref{tab8:fitting value} and Eqn. \ref{eqn12}).

\section{Possible Progenitor Stars of Observed IIP SNe}

\subsection{$SN\;1999em$ and $SN\;1999gi$}

We find that LC profiles of $SN\;1999em$ and $SN\;1999gi$ are identical; their $V$ band and $R$ band are nearly constant until 95 days after the explosion, which suggests these two SNe may originate from two similar progenitor stars (see Figure \ref{fig15:factors to light curve}-(b)). \cite{Utrobin(2007)} and \cite{Berstenetal.(2011)} suggest that $SN\;1999em$ came from a $19\;\Msun$ star with an explosion energy of $1.25\;\foe$.  On the other hand, \cite{Baklanovetal.(2005)} shows that the progenitor of $SN\;1999em$ came from a $15\;\Msun$ star with a metallicity of 0.03 and an explosion energy of $\approx0.6\;\foe$. 
However, our fitting results suggest that the progenitor of  $SN\;1999em$ and $SN\;1999gi$ likely come from a $13\;\Msun$ solar metallicity star with an explosion energy of $1.2\;\foe$. The discrepancies among these models may be caused by mass loss recipes and explosion setups adopted in these studies. Nevertheless, \cite{Smarttetal.(2002)} used pre-explosion images of $SN\;1999em$ to infer its progenitor mass ranging from $11\;-\;13\;\Msun$, consistent with  \cite{Elmhamdietal.(2003)}. Besides, the $M_{\rm Ni}$ observed by \cite{Elmhamdietal.(2003)} is $\sim 0.02\;\Msun$, matching well with our findings. 

\subsection{$SN\;2003gd$}

Although the early plateau-phase data of $SN\;2003gd$ is missing, the rapid change from plateau to linear suggests the SN faints quickly, revealing that the $M_{\rm Ni}$ should be smaller. If the best-fit $M_{\rm Ni}\;=\;0.02\;\Msun$, we infer the progenitor's initial mass should be more than $15\;\Msun$ to align with the observation. The fitting explosion energy is $0.6\;\foe$; if the progenitor mass is lower than we predict, its energy should decrease $0.3\;-\;0.4\;\foe$, a low-explosion-energy SN. However,  \cite{Hendryetal.(2005)} suggests initial mass for $SN\;2003gd$ is $8^{\;+4}_{\;-2}\;\Msun$, which differs from our results. We conjecture that the difference comes from the incomplete data (see Figure \ref{fig15:light curve}-(c)). 

\subsection{$SN\;2004dj$}
Since the progenitor of $SN\;2004dj$ was in a compact star cluster (Sandage 96) in NGC 2403. The age of a cluster can be determined more precisely than that of a single star. Thus, the evolutionary state of the progenitor could be assured by the analysis of the cluster SED (spectral energy distribution) \citep{Vinketal.(2006)}. 

In the study of \cite{Vinketal.(2006)}, the fitting parameters include cluster age, cluster mass, and cluster reddening. The results yielded a "young" population with a cluster age of around 8 million years and an "old" population with a cluster age between 20 and 30 million years separately The progenitor mass of $SN\;2004dj$ in the young population was predicted to $15\;-\;25\;\Msun$ and further discussed in theoretical models by \cite{Kotaketal.(2005)}. These results suggest the progenitor mass of $SN\;2004dj$ $> 15\;\Msun$, which is close to our best-fit model of $13\;\Msun$ (see Figure \ref{fig15:light curve}-(d)). 

\subsection{$SN\;2004et$}

Analyzing $SN\;2004et$ is challenging because the data during plateau to linear transition is unavailable in \cite{Sahuetal.(2006)}. Besides, its best-fit LC suggests the decline of luminosity is rather slow. \cite{Sahuetal.(2006)} suggests a $20\;\Msun$ progenitor for $SN\;2004et$. However, \cite{Lietal.(2005)} analyzed CFHT images of the SN before the explosion and suggested the progenitor as a yellow supergiant of $13\;-\;20\;\Msun$, exceeding our best-fit model of $13\;\Msun$. 

\subsection{$SN\;2005cs$, $SN\;2008in$ and $SN\;2009js$}

$SN\;2005cs$ is a low-luminosity (the peak in $V$ band magnitude > -15), low-energy, Ni-poor CCSN. This faint SN can be modeled by either a core-collapse explosion with a massive star \citep{Zampierietal.(1998)}, or by a less massive progenitor with weak explosion \citep{Chugai&Utrobin(2000)}. To verify that, we have modeled both conditions and inferred that $M_{\rm Ni}$ should be smaller than $0.02\;\Msun$. (see Figure \ref{fig15:light curve}-(f)). However, \cite{Lietal.(2006)} suggested that $SN\;2005cs$ originated from an RSG with an initial mass of $7\;-\;9\;\Msun$ based on the HST/ACS and NICMOS data for M51.

\cite{Gandhietal.(2013)} suggests the very low $M_{\rm Ni}$ of $\sim 0.004\;-\;0.011\;\Msun$ and explosion energy of $0.14\pm0.11\;\foe$ for $SN\;2009js$ that is similar with other low-luminosity events like $SN\;2005cs$ and $SN\;2009N$. Furthermore, they suggested that $SN\;2009js$ originated from a progenitor with $11\pm5\;\Msun$. Such a small $M_{\rm Ni}$ and explosion energy is unusual for SNe IIP, and they may be associated with low-energy SNe, e.g., \cite{Chenetal.(2017)}. 

Our best-fit LCs for $SN\;2008in$ closely match the observation and analysis presented in \cite{Royetal.(2011)}. Our model can explain the photometric characteristics and the plateau duration of approximately 100 days for $SN\;2008in$. The optical LCs, particularly in the $R$ and $I$ bands in $SN\;2008in$, show fluctuations during the plateau phase. The $VRI$ bands of LCs suggest that the change in brightness from mid-plateau to late nebular phase is similar to that of the subluminous $SN\;2005cs$. 
\citep{Gandhietal.(2013)} suggested that $SN\;2005cs$, $SN\;2008in$, and $SN\;2009js$ share characteristics of low luminosity. 

\subsection{$SN\;2009N$ and $SN\;2012A$}

\cite{Taktsetal.(2014)} suggest the progenitor of $SN\;2009N$ came from a $13-13.5\;\Msun$ with an explosion energy of $0.5\;\foe$ and $M_{Ni}\;=\;0.02\;\Msun$ star. Meanwhile, \citet{Maguireetal.(2012)} suggests the progenitor mass of $SN\;2009N$ is $<16\;\Msun$. However, our best fit suggests the progenitor of $20\;\Msun$ with an explosion energy of $0.5\;\foe$.

$SN\;2012A$ is identified as SN IIP with a shorter plateau duration. We estimate its $M_{Ni}\;\approx\;0.02\;\Msun$. \cite{Tomasellaetal.(2013)} suggests the progenitor mass $10.5_{-2}^{+4.5}\;\Msun$ and a double-peaked structure in the $H\alpha$ profile, potentially indicating asymmetric ejection of \Ni. Our model of $15\;\Msun$ cannot fit well due to the fast change in luminosity between the plateau and decay phase. The SN brightness drops by a factor of ten during $96\;-\;106$ days; it can be explained by a massive but Ni-poor SN.

\subsection{$SN\;2012aw$ and $SN\;2012ec$}

$SN\;2012aw$ has been comprehensively studied \citep{Fraseretal.(2012), Kochaneketal.(2012), VanDyketal.(2012), Dall'Oraetal.(2014), Jerkstrandetal.(2014)}. These models suggest the progenitor's mass of $SN\;2012aw$ is $16\;-\;18\;\Msun$, which also accords with our best fit.

\cite{Dall'Oraetal.(2014)} suggests $SN\;2012aw$ containing $\sim 20\;\Msun$ ejecta, explosion energy of $1.5\;\foe$, and \Ni mass of $\sim 0.06\;\Msun$. Nevertheless, our $0.6\;\foe$ models look fainter than $SN\;2012aw$, and another similar model with $1.5\;\foe$ fails to match the observation. We suspect that the explosion energy of $SN\;2012aw$ should be $0.6\;-\;1.5\;\foe$.

Observational data of $SN\;2012ec$ is relatively scarce. \cite{Barbarinoetal.(2015)} used MARCS model SEDs to derive probable progenitor mass for $SN\;2012ec$ should be $14\;-\;22\;\Msun$, also \cite{Jerkstrandetal.(2015)} used spectral synthesis modeling to specify its mass range as $13\;-\;15\;\Msun$. Our best fit shows that this SN may come from a $20\;\Msun$ with a low-energy explosion of $0.6\;foe$, and $0.35\;-\;0.45\;\Msun$ \Ni.

\begin{table}[t]
\hskip-1.5cm
\centering
\resizebox{9cm}{!}{%
\begin{tabular}{|l|ccccc|}
\hline
\multicolumn{1}{|l|}{Progenitor star} & Phrase (day) & $B\;-\;V$ & $T_{\rm eff}$ (K) & {[}Fe/H{]} & $Z_{*}$                        \\ \hline
$SN\;1999em$               & 6            & -0.036                         & 10204         & 0.6942     & 0.06627                        \\
$SN\;1999gi$               & 0.62         & -0.05                          & 10400         & 0.7474     & 0.07490                        \\
$SN\;2003gd$               & 92           & 1.199                          & 4317.9        & -0.0787    & {\color[HTML]{3166FF} 0.01118} \\
$SN\;2004A$                & 34.7         & 0.607                          & 5572.8        & -0.9121    & 0.00164                        \\
$SN\;2004dj$               & 36           & 0.615                          & 5863.7        & 0.128      & 0.01799                        \\
$SN\;2004et$               & 7.7          & 0.23                           & 7650          & 0.1279     & 0.01799                        \\
$SN\;2005cs$               & 3.36         & -0.134                         & 14250         & 2.8709     & {\color[HTML]{FE0000} 9.95416} \\
$SN\;2007od$               & 9.21         & 0.168                          & 8016          & 0.0431     & 0.01480                        \\
$SN\;2008in$               & 6            & 0.24                           & 7425          & -0.2258    & 0.00797                        \\
$SN\;2009bw$               & 5.4          & 0.072                          & 8816.8        & 0.0172     & 0.01394                        \\
$SN\;2009js$               & 3.4          & 0.355                          & 6914.5        & 0.057      & 0.01528                        \\
$SN\;2009N$                & 10.9         & 0.292                          & 7212.3        & -0.0237    & 0.01269                        \\
$SN\;2012A$                & 3            & -0.1                           & 12084.6       & 1.7875     & {\color[HTML]{FE0000} 0.82149} \\
$SN\;2012aw$               & 1.9          & -0.07                          & 10700         & 0.8331     & 0.09124                        \\
$SN\;2012ec$               & 10.22        & 0.18                           & 7950          & 0.0756     & 0.01595                        \\
$SN\;2013ab$               & 5.84         & 0.057                          & 9037.4        & 0.3199     & 0.02799                        \\
$SN\;2013ej$               & 7.28         & -0.017                         & 9938          & 0.609      & 0.05446                        \\ \hline
\end{tabular}%
}
\caption{The prediction of metallicity for each CCSN at the end of stellar evolution. In this table, we first calculate $(B\;-\;V)_{0}$. To get the value more precisely, we choose the data closest to its explosion date ($<10$days). We refer to the empirical value from \cite{Sekiguchi&Fukugita(2000)} and use the way of interpolation to get $T_{\rm eff}$ with $(B\;-\; V)_{0}$. We obtain the value of $\rm [Fe/H]$ with these results and consider it the metallicity when the star is at the end of stellar evolution. However, no data exists in the plateau phase of $SN\;2004A$, $SN\;2004dj$, $SN\;2004et$.  On the other hand, $Z_{*}$ of $SN\;2003gd$ we calculate may be somehow controversial (blue part) since the data we obtained is far from the time of the explosion (nearly 100 days), the SN has been completely split, and its value is meaningless. Also, we guess the reason for $Z_{*}$ of $SN\;2005cs$ and $SN\;2012A$ look unrealistic (red part) because the formula from Eqn. \ref{eqn11} may not be applicable for high-temperature surface ($>10^{4}\;\K$).}
\label{tab8:fitting value}
\end{table}

\section{Effects of Stellar Parameters on Light Curves}

\begin{figure*}[ht]
    \centering
    \subfigure[]{
        \includegraphics[scale=0.35]{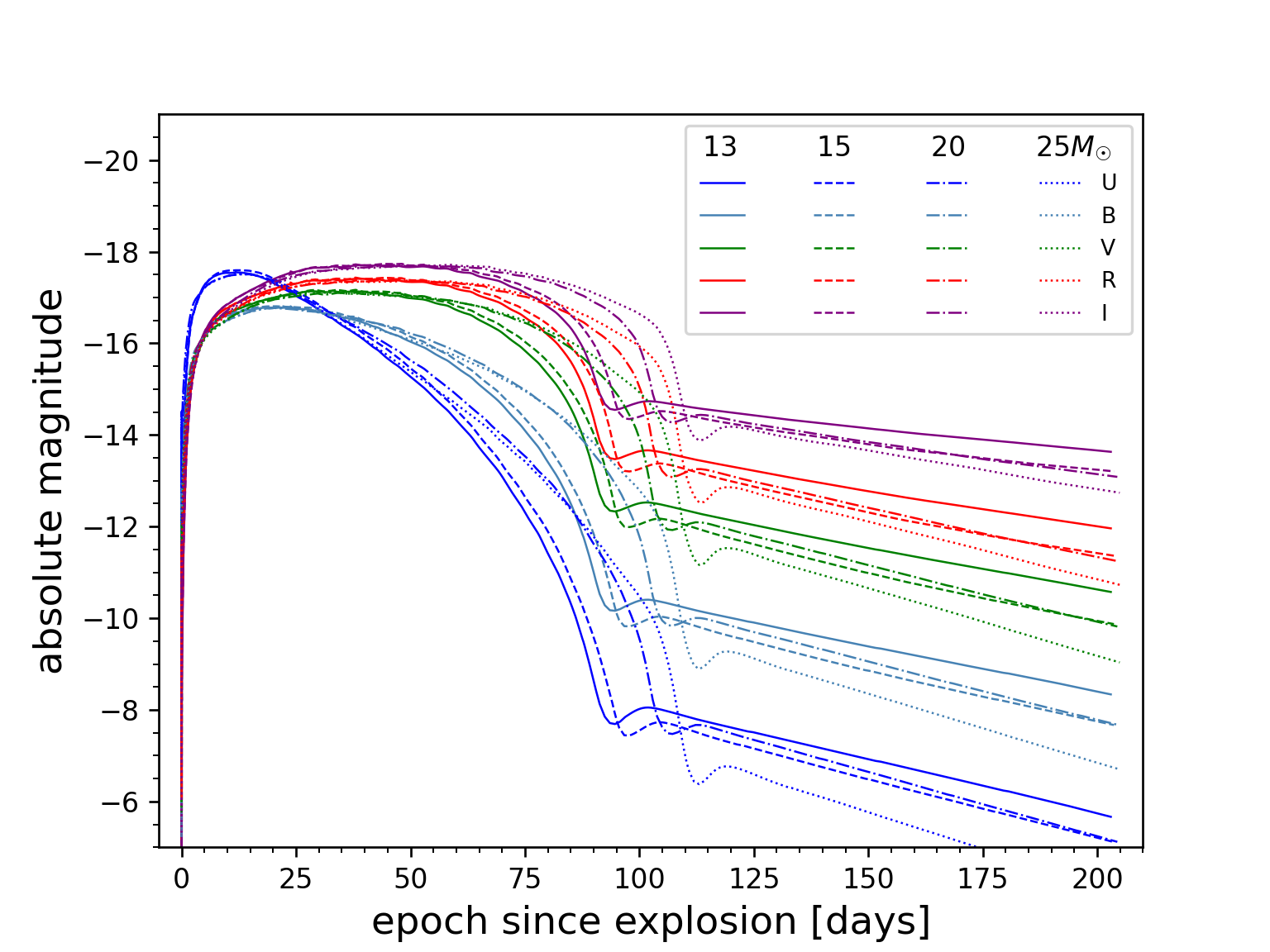}
        }
    \subfigure[]{
        \includegraphics[scale=0.35]{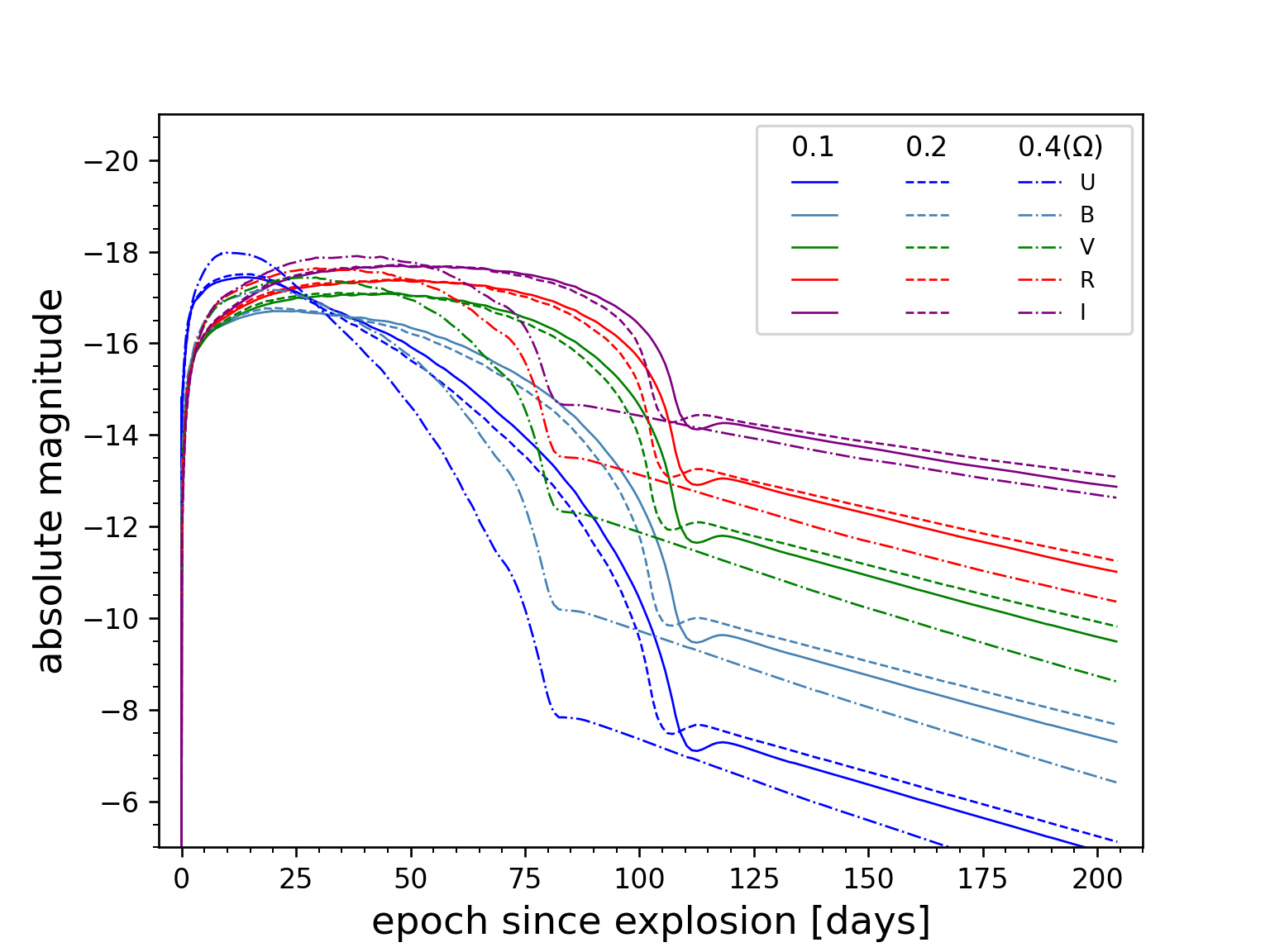}
        }
    \subfigure[]{
        \includegraphics[scale=0.35]{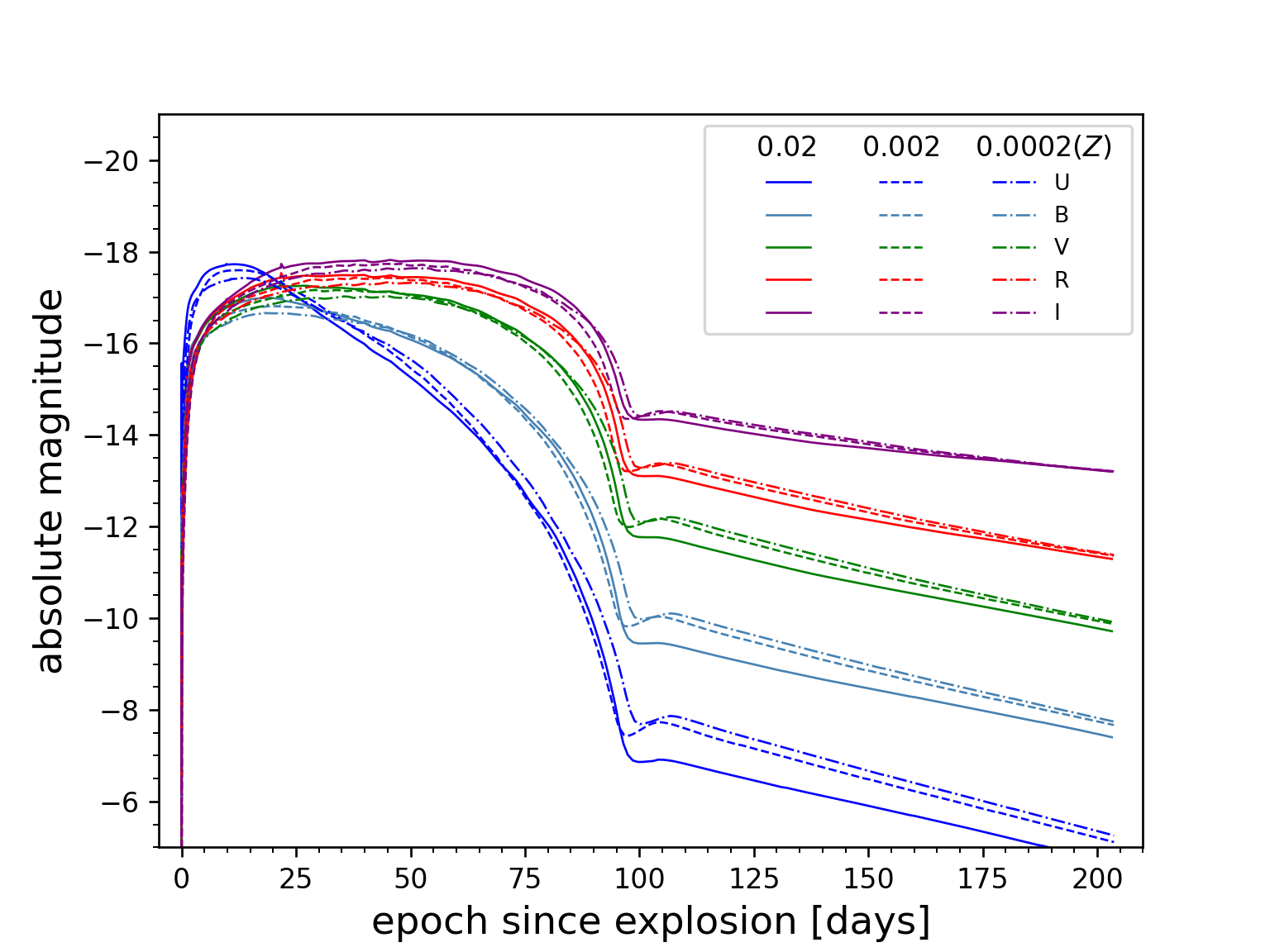}
        }
    \subfigure[]{
        \includegraphics[scale=0.35]{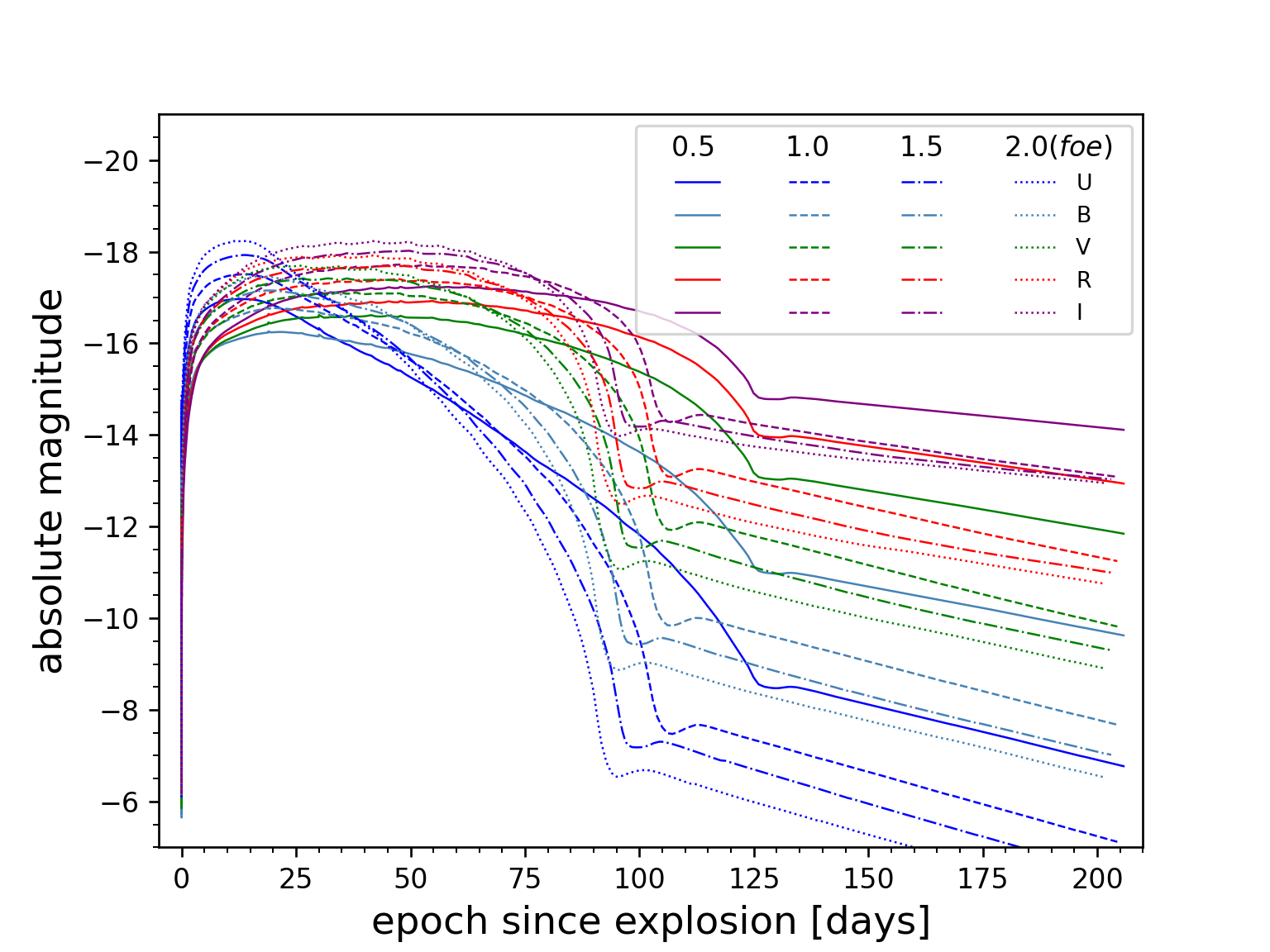}
        }
    \subfigure[]{
        \includegraphics[scale=0.35]{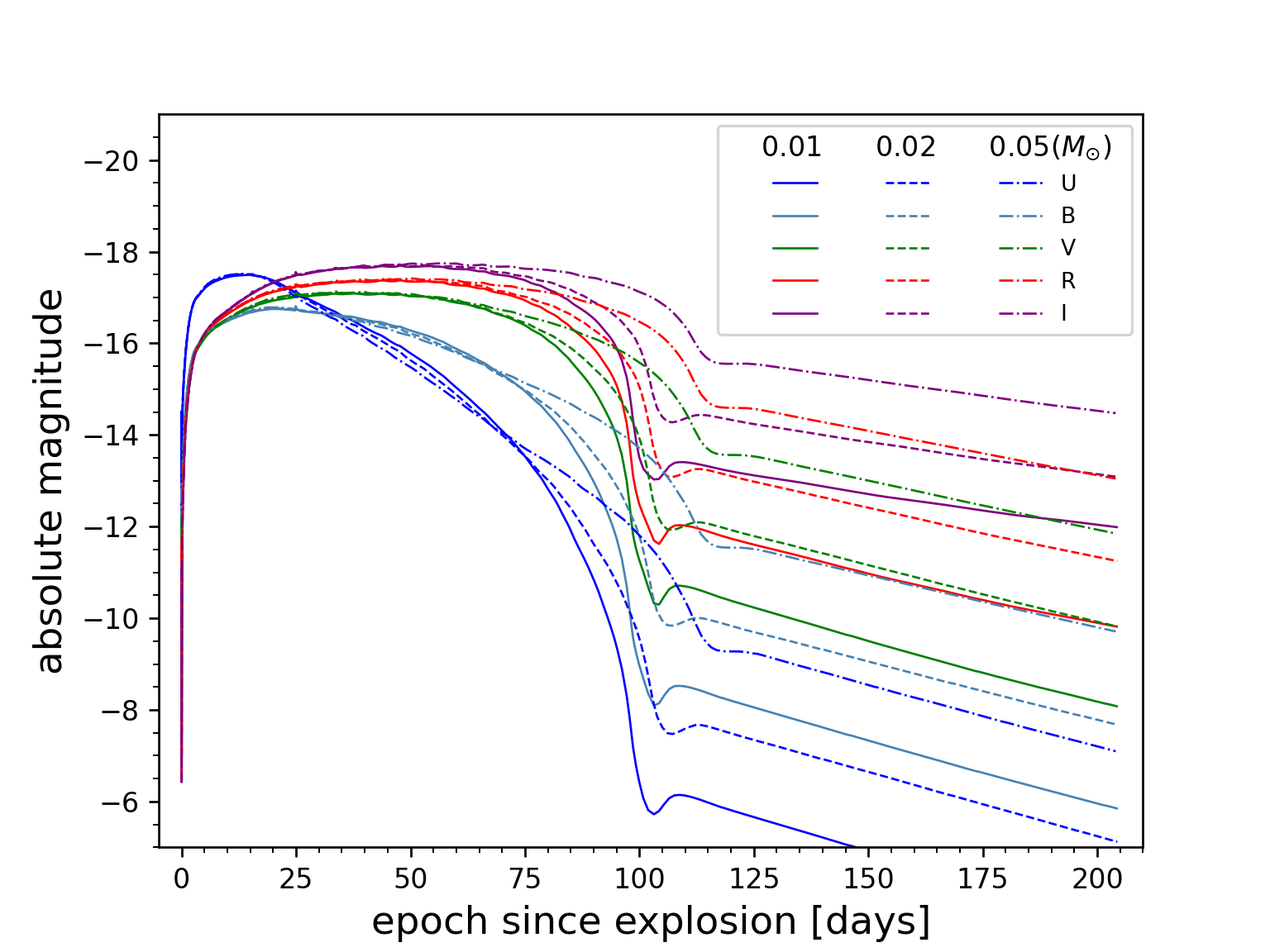}
        }
    \caption{The LC formation considered with various independent variables. (a) models of different progenitor's masses during the PMS with controlled factors  $\Omega\;=\;0.2$ / $Z\;=\;0.002$ / $M_{\rm Ni}\;=\;0.02\;\Msun$ / $1.2\;\foe$. (b) models of different rotational rates during the PMS with controlled factors $20\;\Msun$ / $Z\;=\;0.002$ / $M_{\rm Ni}\;=\;0.02\;\Msun$ / $1.2\;\foe$. (c) models of different metallicities during the PMS  with controlled factors $15\;\Msun$ / $\Omega\;=\;0.2$ / $M_{\rm Ni}\;=\;0.02\;\Msun$ / $1.2\;\foe$. (d) models of injecting various explosion energies at the end of evolution with controlled factors $20\;\Msun$ / $\Omega\;=\;0.2$ / $Z\;=\;0.002$ / $M_{\rm Ni}\;=\;0.02\;\Msun$. (e) After initialing explosion, we inject models with different $M_{\rm Ni}$ at the SN's center with controlled factors $20\;\Msun$ / $\Omega\;=\;0.2$ / $Z\;=\;0.002$ / $1.2\;\foe$.}
    \label{fig15:factors to light curve}
\end{figure*}

\subsection{Mass}

In Figure \ref{fig15:factors to light curve}-(a), the curve of the first peak of the LCs looks similar since we set the same input explosion energy in all models. The difference appears at the duration of the plateau. Although the increasing stellar mass also enhances its mass loss, the more massive stars can still retain more \Hy. For the $13\;\Msun$ star, it remains $5.3026\;\Msun$ hydrogen before the star goes through the explosive stage, while the $25\;\Msun$ star remains $6.6508\;\Msun$. The difference in their hydrogen mass before an explosion can explain the increasing duration of the plateau phase \citep{Litvinova&Nadezhin(1985), Popov(1993)}.

Additionally, the composition of the ejecta, including the presence of hydrogen and heavy elements, influences the opacity and energy transport within the SN, impacting the LC's shape. More specifically, the envelope structure of a massive star can affect the interaction between the shockwave and the envelope material. An extensive envelope can lead to more prolonged interactions and a slower decline in luminosity during the decay phase of the LC \citep{Hillier&Dessart(2019)}. 

On the other hand, SN's $M$ from the more massive star declines faster in the decay phase of the LC because we inject the same amount of $M_{\rm Ni}$ in the SN's center in models of various masses. For more massive models, the decay energy of \Ni\ and \Co\ is used to expand the ejecta instead of emitting radiation. 

\subsection{Rotation}

The stellar rotation creates various kinds of mixing and affects their internal profiles of density, temperature, and chemical composition. This also influences the stellar radius, luminosity, and core properties during the pre-supernova phase. Rapid rotation can increase the angular momentum of the core, affecting the properties of the collapsing core. These dynamics can influence the initial shock breakout and the subsequent propagation of shockwave. Also, it enhances convective mixing processes within the SN. This can lead to dredging up heavy elements from the core to the surface layers and vice versa, leading to brighter LCs (see Figure \ref{fig15:factors to light curve}-(b)). In addition, rapidly rotating stars can experience enhanced mass loss due to the rotation-driven wind. The amount and rate of mass loss can affect the total amount of ejecta and its composition, both of which influence SN's brightness duration.

\subsection{Metallicity}

Higher metallicity generally leads to higher opacity in the stellar envelope due to the increased presence of heavy elements. Opacity affects how radiation transports in the SN ejecta, affecting the overall luminosity and shape of the LC. High-metallicity stars typically produce thicker oxygen and silicon-burning shells. During the explosion, the interaction between the shockwave and these shells can lead to additional emission and absorption features in the LC.

The higher metallicity model appears more like a broad peak compared to a plateau in low-metallicity models due to a strong mass loss of the hydrogen envelope (see Figure \ref{fig15:factors to light curve}-(c)). Furthermore, metallicity can also affect nuclear burning and alter elemental yields, affecting the LCs.

\subsection{Explosion Energy}

The explosion energy directly determines the initial luminosity of SNe IIP. A higher explosion energy can produce a strong shock that deposits more energy in the stellar envelope and produces a brighter LC. (see Figure \ref{fig15:factors to light curve}-(d)). Meanwhile, the explosion energy determines the expanding velocity of SN ejecta. Higher energy accelerates the ejecta to higher velocity, shrinking the photon diffusion time and resulting in a shorter plateau duration.

The explosion energy affects the nucleosynthetic yields during the SN explosion. Higher energy explosions produce stronger explosive burning, producing a broader range of elements. The resulting composition of the ejecta influences radiation processes, which are reflected in the decay-phase of LC.

\subsection{$^{56}Ni$ Production}

\Ni\ first decays to \Co\ and then to stable \Fe. This decay chain produces gamma-ray photons that deposit their energy into the surrounding ejecta \citep{Srinivasan(2014), Hotokezakaetal.(2016)}, heating it and powering the luminosity of the SN in the decay phase \citep{Chenetal.(2020)}. Additionally, the timescale of the luminosity decline is determined by the radioactive decay of \Co. A larger $M_{\rm Ni}$ results in a slower decline, leading to a more extended plateau phase in the LC (see Figure \ref{fig15:factors to light curve}-(e)).

\section{Discussion}
\subsection{Evolution of progenitor stars}
Based on our simulations, when $Z$ is high, the star's evolution is accelerated with an enhanced mass loss. Still, the mass loss can be dominated by rotation for low-metallicity stars. As $Z$ goes up, \Ox\ and \Si\ are more likely to be consumed rapidly in the core, and as $\Omega$ increases, the size of carbon and oxygen core also shrinks. A more massive iron core can form under the presence of $\Omega$. Rapidly rotating models dredge significant heavy elements to the surface to substantially raise surface metallicity, enhancing the mass loss.

\subsection{Discrepancy between observation and model}
Although our plateau phase of LCs are roughly consistent with observation, some deviations remain due to the uncertainties of various stellar parameters of $Z$, $\Omega$, mass on the evolution, and explosion parameters of $M_{\rm Ni}$, mixing, and explosion energy. So far, we have determined that massive star explosions lead to SNe IIP  and identified the physical properties of their progenitor stars. In our future work, we will obtain more accurate values for the exact ejecta mass, metallicity, rotational rates, explosion energy, and elemental yields by improving our current simulations.

\subsection{Modeling Issues}
\label{sec:modeling}
\subsubsection{Before Core-collapse}

Some models fail to reach numerical convergence and crash. This issue may be resolved by increasing the interaction numbers at the expense of long run time.  However, in the case of $13\;\Msun\;/\;\Omega\;=\;0.4\;/\;Z\;=\;0.002$, the star degenerates during its silicon burning and fails to evolve further. Even though we have tried to adjust the wind and overshooting parameters, this model remains unsuccessful.

For models with a strong rotation of $\Omega \geq 0.5$, strong mixing makes the hydrogen envelope burning to reach the surface, creating a gigantic He-rich star before getting the RSG branch. Such a star could continue evolving to core collapse and become a Type Ib/c SN \citep{Yoon&Langer(2005), Rileyetal.(2021)}. If we want an RSG with low metallicity and high rotation speed, we must decrease the wind efficiency and overshoot to retain more of the original hydrogen envelope.

\subsubsection{Collapse and Explosion}

Evolution can encounter challenges during the core collapse phase due to a significant spike in entropy at the outset caused by the injection of explosive energy into a small region. This spike often signifies mixing that cannot be adequately modeled in 1D simulations, leading to numerical instability and crashes. Adjustments to the infall radius and explosion parameters are necessary to enhance stability.

Using a thermal bomb to simulate the explosion of a CCSN is an artificial approach, as it does not capture the physical processes of neutrino heating and shock revival. However, the choice of bomb parameters slightly affect the dynamics of SN ejecta and the resulting light curves \citep{Paxtonetal.(2018)}. 

\section{Conclusion}

We present a comprehensive study on the origins of SNe IIP, shedding light on the physical properties of their progenitor stars by examining the impact of multiple stellar factors such as metallicity, mass, rotation, and explosion. Through fitting over 15 observed SNe IIP; most of them have best-fit models to explain their physical origins. We can conclude that most progenitors stem from moderate-massed MS between $13\;-\;20\;\Msun$ with $\Omega$ between 0.1 to 0.4. For explosion, over half of SN carries a low explosion energy ($0.6\;\foe$) and $0.02\;\Msun$ \Ni. The LCs from our models roughly align with observations but with minor discrepancies. Issues may stem from an amount of $M_{\rm Ni}$, various stellar parameters ($Z$, $\Omega$, yields), and micro-physics of stellar evolution, such as nuclear burning and explosion mechanisms.

Our simulations demonstrate that high-metallicity stars lead to accelerated evolution and enhanced mass loss, while rotation ($\Omega$) dominates at low metallicity. Higher $Z$ and $\Omega$ accelerate the burning of \Ox\ and \Si, affecting the final yields of SNe. Rapidly rotating models induce strong mixing that dredges the heavy elements to the stellar surface and contributes to mass loss via radiative-driven wind. Various stellar factors intricately shape the LC of a SN.

For the observational signatures of SNe IIP, more massive progenitors result in longer plateaus of their LCs due to massive ejecta, while rapidly rotating stars exhibit enhanced convective mixing and mass loss, potentially affecting the LCs. The higher metallicity model forms more like a broad peak feature than a plateau in LC due to the reduced hydrogen envelope through mass loss. Explosion energy determines the early luminosity and plateau duration, with higher energies yielding more luminous and shorter plateaus due to faster ejecta. \Ni\ production influences late-time luminosity in which larger \Ni\ mass extends the plateau duration.

Stellar parameters impact the SN LCs through intricate dynamics of stellar explosions, nucleosynthesis, and radiation transfer. Comprehending these interplay processes offers insight into the physics of SNe IIP and their progenitor stars, thereby enhancing our understanding of observed supernovae phenomena.

\section{Acknowledgement}

This research is supported by the National Science and Technology Council, Taiwan, under grant No. MOST 110-2112-M-001-068-MY3 and the Academia Sinica, Taiwan, under a career development award under grant No. AS-CDA-111-M04.  Our computing resources were supported by the National Energy Research Scientific Computing Center (NERSC), a U.S. Department of Energy Office of Science User Facility operated under Contract No. DE-AC02-05CH11231 and the TIARA Cluster at the Academia Sinica Institute of Astronomy and Astrophysics (ASIAA).

\software{\MESA\ \citep{Paxtonetal.(2011)}, \STELLA\ \citep{Blinnikovetal.(1998)}, Python}

\end{CJK*}

\end{document}